%% file: main_double_final.tex
\documentclass{IEEEtaes}

\usepackage{color,array,amsthm}
\usepackage{graphicx}
\usepackage{cite}
\usepackage{amsmath,amsfonts}
\usepackage{graphicx}
\usepackage{textcomp}
\usepackage{xcolor}
\usepackage{booktabs}
\usepackage{multirow}
\usepackage{lipsum}
\usepackage{subfigure}
\usepackage{stfloats}% <-- added

\usepackage[capitalise,noabbrev,nameinlink]{cleveref}
\crefname{appfig}{Appendix Figure}{Appendix Figures}
\crefname{apptab}{Appendix Table}{Appendix Tables}
\crefname{appsec}{Appendix Section}{Appendix Sections}

\Crefname{section}{Section}{Sections}
\crefname{section}{Section}{Sections}
\Crefname{subsection}{Subsection}{Subsections}
\crefname{subsection}{Subsection}{Subsections}

\usepackage[makeroom]{cancel}
\usepackage{bbm}
\usepackage{amssymb,bm}  % assumes amsmath package installed

\usepackage[acronyms,nonumberlist,nopostdot,nomain,nogroupskip,acronymlists={hidden}]{glossaries} 
\newglossary[algh]{hidden}{acrh}{acnh}{Hidden Acronyms}

\usepackage{algorithmicx}
\usepackage[noend]{algpseudocode}
\algrenewcommand\algorithmicindent{1em} 

\usepackage{algorithm2e}

\jvol{XX}
\jnum{XX}
\jmonth{XXXXX}
\paper{1234567}
\pubyear{2024}
\doiinfo{TAES.2024.Doi Number}

\usepackage{caption}

\newcommand{\norm}[1]{\lVert {#1} \rVert}
\newcommand{\chib}{\bm{\chi}}
\newcommand{\uvec}{\bm{u}}

\newcommand{\vvec}{\bm{v}}
\newcommand{\bigvvec}{\bm{V}}
\newcommand{\Jb}{\bm{J}}
\newcommand{\Db}{\bm{D}}
\newcommand{\Wb}{\bm{W}}
\newcommand{\Ab}{\bm{A}}
\newcommand{\Ib}{\bm{I}}

\input{acronyms}

\begin{document}

\title{Continuously Optimizing Radar Placement with Model Predictive Path Integrals}
% \title{Optimal Bayesian Target Recognition for Unmanned Aerial Vehicles Based on Radar Cross Section} 

\author{Michael Potter}
\member{Student Member, IEEE}
\affil{Northeastern University, Boston, MA 02115, USA} 

\author{Shuo Tang}
\member{Student Member, IEEE}
\affil{Northeastern University, Boston, MA 02115, USA} 

\author{Paul Ghanem}
\member{Student Member, IEEE}
\affil{Northeastern University, Boston, MA 02115, USA} 

\author{Milica Stojanovic}
\member{Fellow, IEEE}
\affil{Northeastern University, Boston, MA 02115, USA} 

\author{Pau Closas}
\member{Senior Member, IEEE}
\affil{Northeastern University, Boston, MA 02115, USA} 

\author{Murat Akcakaya}
\member{Senior Member, IEEE}
\affil{University of Pittsburgh, Pittsburgh, PA 15260, USA}

\author{Ben Wright}
%\member{Member, IEEE}
\affil{Kostas Research Institute, Burlington, MA, USA} 

\author{Marius Necsoiu}
\member{Member, IEEE}
\affil{DEVCOM ARL, San Antonio, TX 78204, USA} 

\author{Deniz Erdo\u{g}mu\c{s}}
\member{Senior Member, IEEE}
\affil{Northeastern University, Boston, MA 02115, USA} 

\author{Michael Everett\textsuperscript{*}}
\member{Member, IEEE}
\affil{Northeastern University, Boston, MA 02115, USA} 
%% \author{FOURTH D. AUTHOR}
%% \affil{University of Colorado, Colorado, USA}

\author{Tales Imbiriba\textsuperscript{*}}
\member{Member, IEEE}
\affil{University of Massachusetts Boston, MA 02125, USA}

\receiveddate{\tiny Manuscript received May 29, 2024; revised December 18, 2024; accepted December 30, 2024.\\
* denotes equal contribution of authors.
Research was sponsored by the Army Research Laboratory and was accomplished under Cooperative Agreement Number W911NF-23-2-0014. The views and conclusions contained in this document are those of the authors and should not be interpreted as representing the official policies, either expressed or implied, of the Army Research Laboratory or the U.S. Government. The U.S. Government is authorized to reproduce and distribute reprints for Government purposes notwithstanding any copyright notation herein.
}

\markboth{POTTER ET AL.}{Continuously Optimizing Radar Placement with Model Predictive Path Integrals}
\maketitle
\begin{abstract} 
Continuously optimizing sensor placement is essential for precise target localization in various military and civilian applications. While information theory has shown promise in optimizing sensor placement, many studies oversimplify sensor measurement models or neglect dynamic constraints of mobile sensors. To address these challenges, we employ a range measurement model that incorporates radar parameters and radar-target distance, coupled with \gls{mppi} control to manage complex environmental obstacles and dynamic constraints. We compare the proposed approach against stationary radars or simplified range measurement models based on the \gls{rmse} of the \gls{ckf} estimator for the targets' state. Additionally, we visualize the evolving geometry of radars and targets over time, highlighting areas of highest measurement information gain, demonstrating the strengths of the approach. The proposed strategy outperforms stationary radars and simplified range measurement models in target localization, achieving a 38-74\% reduction in mean \gls{rmse} and a 33-79\% reduction in the upper tail of the 90\% \gls{hdi} over 500 \gls{mc} trials across all time steps.
Code available at \href{https://github.com/mlpotter/OptimalTrackingAndControl}{https://github.com/mlpotter/OptimalTrackingAndControl}.
\end{abstract}

\vspace{-1.5em}

\glsresetall

\input{introduction/introduction_draft_r1}
\vspace{-1.5em}
\input{relatedwork/relatedwork_draft_r1}
\section{Preliminaries}
\label{sec:preliminaries}

We consider $N$ radars and $M$ targets. The state for radar $n$ at time step $k$ is denoted as $\chib^{R_n}_k = [x,y,z,\theta,v,\omega]$, while for target $m$, it is $\chib^{T_m}_k = [x,y,z,\dot{x},\dot{y},\dot{z}]$, where $\theta$ is the radar's heading angle with respect to the $x$-axis of the radar coordinate frame, $v$ is the heading velocity, and $\omega$ is the angular velocity. The positions for radar $n$ and target $m$ are defined as $\chib^{R_n}_{xyz} = [x,y,z]$ and $\chib^{T_m}_{xyz}=[x,y,z]$, respectively. The time step $k$ has a duration of $\Delta t$. We denote the Hadamard product (elementwise multiplication) as $\odot$, and $\otimes$ is the kronecker product. Additionally, we use the notation $\Delta_{x}^{y}=\nabla_{x} \nabla_{y}^T$ for the Hessian operator.

To highlight the focus of our paper, we make the following simplifying assumptions to our analysis:

\begin{enumerate}
    \item \textit{Perfect data association and target detection}. We assume that the process of selecting which measurement(s) to incorporate into the target state estimator, is perfect, and the radars constantly track the targets. 
    \item \textit{Doppler shift can be neglected}. For slow relative velocity $\nu$ between the radar and target, when the Doppler shift is much smaller than the effective bandwidth of the return signal, $\frac{f_{d}}{\beta} \ll 1$, the Doppler shift $f_d = 2 \times \frac{f_c}{c} \left ( \frac{ (\chib^T_{xyz} - \chib^R_{xyz}) ^T}{ |\chib^T_{xyz} - \chib^R_{xyz}|} \right) \nu$ \cite{shames2013doppler} may be disregarded in the radar return signal. $f_c$ is the carrier frequency and $c$ is the speed of light.
    \item \textit{Radars have zero altitude}. The targets move in $R^3$, but the radars move on a plane at ${z=0}$. This is a common assumption that radars are attached to ground vehicles.
    \item \textit{We assume targets follow a constant velocity model with acceleration process noise and therefore do not account for drastic target maneuvers in our model.}
\end{enumerate}

\begin{figure}[h!]
    \centering
    \includegraphics[width=7cm]{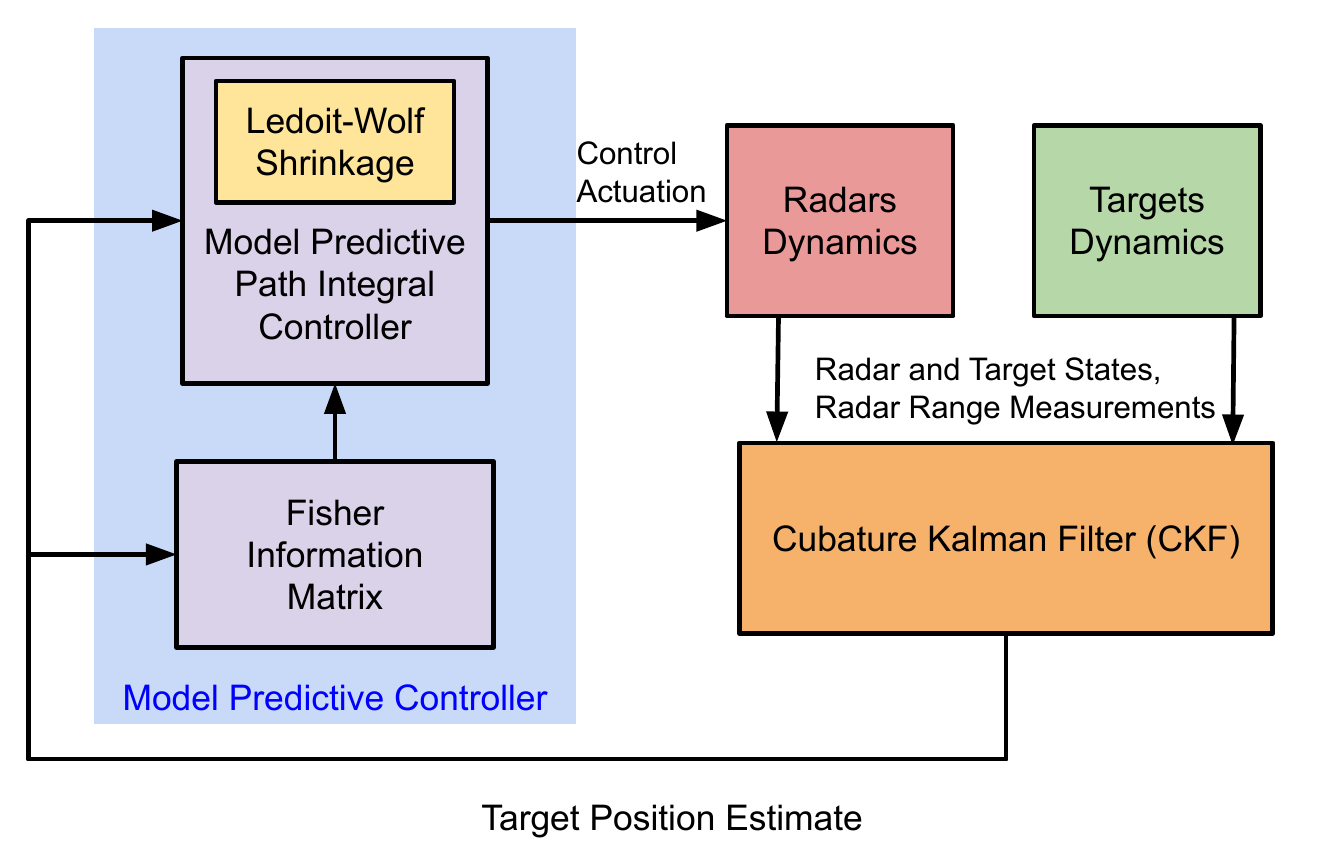}
    \caption{Block Diagram of proposed approach.}
    \label{fig:system_diagram}
    \vspace{-3em}
\end{figure}

With these assumptions, the following sections detail each computation block needed to continuously optimize radar placement (flow diagram depicted in \cref{fig:system_diagram}). We first discuss the more realistic range measurement model based on \cite{godrich2010target} compared to previous works such as \cite{crasta2018multiple,hung2020range,bishop2010optimality}.

\input{methodology/methodology_draft_r1}

\vspace{-1em}
\section{EXPERIMENTS}
\label{sec:experiment}
We examine three scenarios: 1. an ``underdetermined'' system, with fewer radars than targets, 2. an ``overdetermined'' system, with more radars than targets, and 3. an ``full'' system, with equal amount of radar and targets. We highlight not only the poorer performance of a misspecified measurement model (\gls{ddr} versus \gls{ccr} measurement model) but also the pathologies in radar movement caused by the controller using a \gls{fim}-based objective specified from the \gls{ccr} measurement model. The proposed approach outperforms methods employing stationary radars and simplified range measurement models in target localization, achieving a reduction in mean \gls{rmse} and the upper tail of the 90\% \gls{hdi} by 38-74\% and 33-79\% respectively, across 500 \gls{mc} trials at all time steps. 

\vspace{-1em}
\subsection{Simulation}
The hyperparameters for each experiment for the \gls{mppi} control, the \gls{ais}, the radar configurations , and the \gls{mpc} objective is found in \cref{tab:mpc_config,tab:mppi_config,tab:radar_config}. 

\begin{table}[h!]
\centering
\begin{tabular}{|l|l|}
\hline
\multicolumn{1}{|c|}{\textbf{Radar Parameter}} & \textbf{Value} \\ \hline
Carrier Frequency   & $1\times 10^8$             \\ \hline
Radar Loss          & 1                    \\ \hline
Transmitter Gain    & 200                        \\ \hline
Receiver Gain       & 200                        \\ \hline
Radius (for SNR)    & 500 [m]                       \\ \hline
Radar Cross Section & 1 [m\textsuperscript{2}]                         \\ \hline
Power Transmit      & 1000 [W]                          \\ \hline
SNR      & -20 [dB]                          \\ \hline
\end{tabular}
\caption{The Radar Equation configuration settings for each experiment}
\label{tab:radar_config}
\end{table}

\begin{table}[h!]
\centering
\begin{tabular}{|l|l|}
\hline
\multicolumn{1}{|c|}{\textbf{MPPI Parameters}} & \textbf{Value} \\ \hline
$u_{a}$ std   & 25 [m/s\textsuperscript{2}]          \\ \hline
$u_{\dot{w}}$  std       & $45^\circ$                    \\ \hline
Horizon    & 15                        \\ \hline
Number of Sample Trajectories      & 200                        \\ \hline
Number of MPPI Sub-iterations & 5                         \\ \hline
Temperature  & 0.1             \\ \hline
Elite Threshold   & 0.9                        \\ \hline
AIS  Method                 & Cross Entropy                        \\ \hline
Covariance Shrinkage Method       & Ledoit-Wolf Shrinkage                        \\ \hline
Radar Cross Section & 1                          \\ \hline
\end{tabular}
\caption{The MPPI settings for each experiment}
\label{tab:mppi_config}
\end{table}

\begin{table}[h!]
\centering
\begin{tabular}{|l|l|}
\hline
\multicolumn{1}{|c|}{\textbf{MPC Parameter}} & \textbf{Value} \\ \hline
$\gamma$      & $0.95$             \\ \hline
$R2T$         & 125 [m]                    \\ \hline
$R2R$         & 10 [m]                          \\ \hline
$\alpha_1$     & 500                        \\ \hline
$\alpha_2$     & 1000                          \\ \hline
\end{tabular}
\caption{The MPC  configuration settings for each experiment}
\label{tab:mpc_config}
\end{table}

We assume that the radars move on the ground plane (z=0), while the drones move in free space by the constant velocity kinematic model (\cref{eqn:dynamics_single_target}) with acceleration noise constant $\sigma_W=\sqrt{10}$. For each experiment , we conduct 500 \gls{mc} trials, where every \gls{mc} trial is 600 time steps (1 minute in real time). The initial heading velocity and acceleration of the radars are both set to 0, and their initial positions in the XY-plane are uniformly sampled within a square with edge sizes 800 [m], centered at the origin (0 [m], 0 [m]). Although the targets may start in the same position for each \gls{mc} trial, the random acceleration noise in the transition model (\cref{eqn:transition_model}) leads to significant variations in the trajectories of the targets across trials.  The observed range measurements at each time step are simulated using the more realistic range measurement model presented in \cref{eqn:range_measurement}.

We compare our approach quantitatively and qualitatively with stationary radars. Additionally, using the \gls{mppi} controller for mobile radars, we compare the improved measurement model \cite{godrich2010target} to the \gls{ccr} measurement model used in \cite{jiang2015optimal,bishop2010optimality,crasta2018multiple,hung2020range}. This comparison not only empirically demonstrates the suboptimality of the constant covariance model as range measurements become noisier with increasing distance between the radar and the target (due to model mispecification), but also highlights the pitfalls of a controller using an objective based on a \gls{ccr} measurement model. In the \gls{ccr} measurement model, we use an ``optimistic'' diagonal covariance matrix with elements $R2T^4*\Gamma$ meters, where $R2T$ is the radar-target distance constraint.

For experimental purposes, we maintain constant noise power for the \gls{awgn} in \cref{eqn:return_signal}. We establish the desired \gls{snr} level at a specified distance using the formula:
\begin{align}
\sigma_a^2 = \frac{M \tilde{P_r}}{10^{SNR / 10}}
\end{align}
Here, $\tilde{P}_r$ represents the radar received power at the specified distance 500 [m]. For all experiments we set \gls{snr} = -20 [dB].

We note that the \gls{mppi} controller was written in JAX, which on a NVIDIA GeForce RTX 2080 Ti runs at 26-27 Hz (realtime), despite not fully optimizing our code.

\vspace{-1em}
\subsection{Improved Target Localization}
In over 500 \gls{mc} trials, we empirically show that our approach exhibits significantly lower \gls{rmse} for all time steps (see \cref{fig:rmse_compare} (a,b,d,e,g,h)) compared to both the stationary radar approach and the mobile radar approach with constant range measurement. Additionally, the 90\% \gls{hdi} interval in localization performance for the \gls{ccr} measurement model is significantly larger compared to the \gls{ddr} measurement model presented in \cite{godrich2010target} (\cref{fig:rmse_compare} (c,f,i)). The \gls{pfim}-based objective for both range measurement models has slightly higher targets' state estimate \gls{rmse} compared to the \gls{sfim}-based objective. 

However, Bayesian methods are notorious for overfitting, attributed to ensembling with model mispecification and the assumption of a single true model generating the data \cite{6033566,domingos2000bayesian}. We further examine the performance of the proposed approach inn the three simulation scenarios in the next subsections.

\subsubsection{3 Mobile Radars and 4 Targets}
We initialize the targets such that all target directional velocities  point towards the upper right (northeast) in the world coordinate frame, and begin around the same position (see \cref{fig:noise_compare}):
\begin{align*}
    \chi^{T_{1:M}}_0 = \begin{bmatrix}
        0 & 0 & 55 & 20 & 10 & 0 \\
        15.4 & 15.32 & 70 & 15 & 20 & 0 \\
        10 & 10 & 55 & 17 & 19 & 0 \\
        20 & 20 & 45 & 6 & 8 & 0
    \end{bmatrix}
\end{align*}
where the first 3 columns are the 3D xyz coordinates, and the last 3 columns are the directional velocities.

Therefore, with fewer radars than targets, we anticipate a circular formation of radars near the targets when continuously optimizing radar placement, aiming to maintain nearest feasible proximity to the targets.
%maintaining a surrounding radar-target geometry.  

\begin{figure}[h!]
    \vspace{-1em}
    \centering
    \includegraphics[width=0.98\columnwidth]{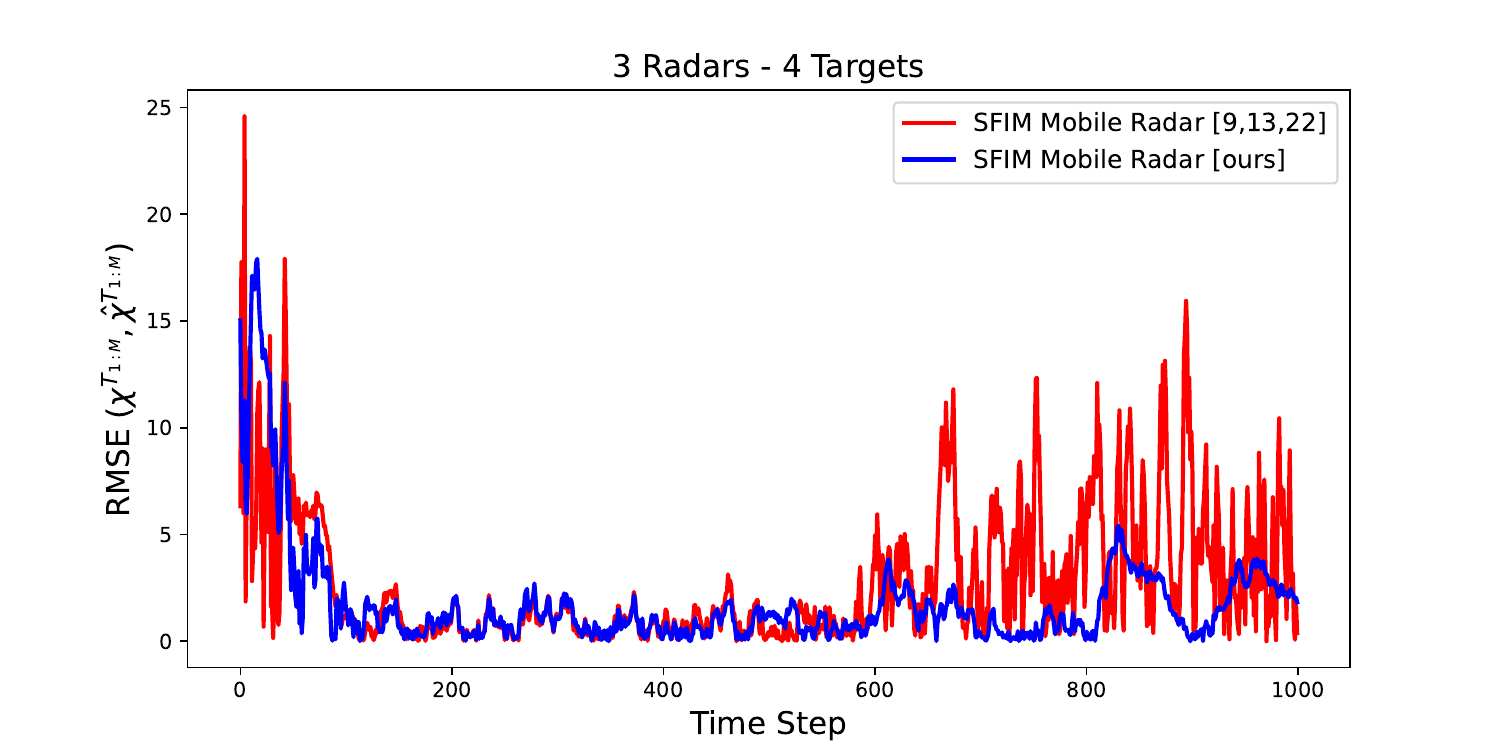}
    \caption{\gls{rmse} for single simulation realization in \cref{fig:noise_compare}}
    \label{fig:mse_compare1}
    \vspace{-1em}
\end{figure}

To illustrate the shortcomings of a controller using a \gls{ccr} measurement model-based objective for radar placement, we showcase four time steps of a \gls{mc} simulation realization, as depicted in \cref{fig:noise_compare}. For our proposed approach, at the beginning of the simulation, radars move towards targets, forming a circular arrangement within the nearest feasible proximity while adhering to radar-target constraints. As targets disperse, radars maintain this formation, but increased range measurement noise prompts a tighter circle, resulting in most targets being slightly outside the radar formation (left subfigure in \cref{fig:noise_compare}). In contrast, approaches with distance-independent noise maintain optimal angular spacing regardless of radar-target distance, often moving very far from the targets (\cref{fig:noise_compare} (b,d,f,h)) and causing high localization errors (\cref{fig:rmse_compare} (a-i)). Counterintuitively, to achieve the optimal angle, mobile radars using the \gls{ccr} measurement model move away from the targets, resulting in larger localization errors (\cref{fig:noise_compare}(b,f)). The target localization error over 1000 time steps for the extended realized \gls{mc} simulation in \cref{fig:noise_compare} is depicted in \cref{fig:mse_compare1}.
\subsubsection{6 Mobile Radars and 3 Targets} 
We initialize each target's velocity direction to be different, with target 1 heading towards the upper right, target 2 towards the lower right, and target 3 towards the lower left:
\begin{align*}
    \chi^{T_{1:M}}_0 = \begin{bmatrix}
        0 & 0 & 70 & 25 & 20 & 0 \\
        -100.4 & -30.32 & 45 & 20 & -10 & 0 \\
        30 & 30 & 80 & -10 & -10 & 0 \\
    \end{bmatrix}
\end{align*}
where the first 3 columns are the 3D xyz coordinates, and the last 3 columns are the directional velocities.

The initial positions of the targets are depicted in \cref{appfig:noise_compare2}.  Since the targets' are spreading out over an open space, we hypothesize that the radars will send atleast 1 radar to follow each target, while keeping a reasonable distance to all other targets. 

\begin{figure}[h!]
    \vspace{-1em}
    \centering
    \includegraphics[width=0.98\columnwidth]{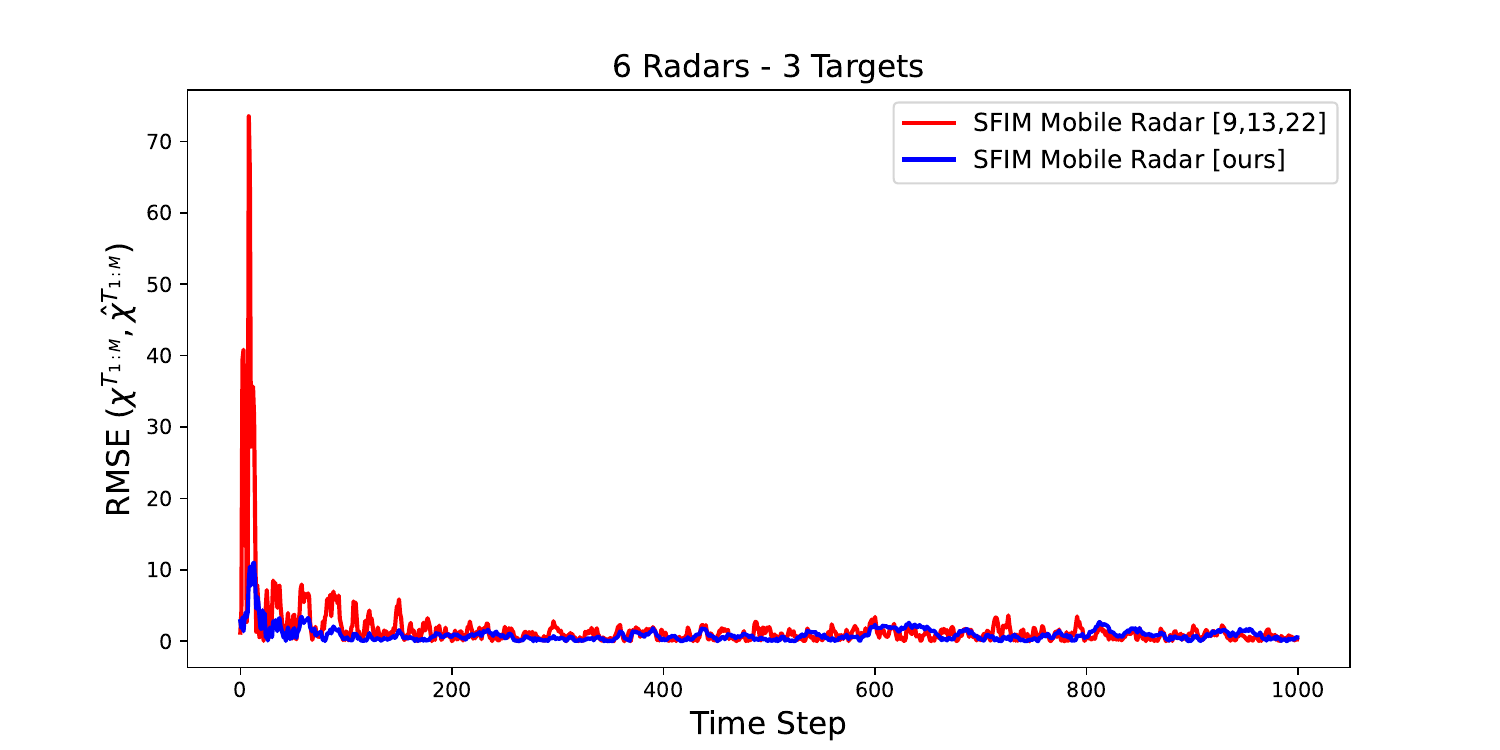}
    \caption{\gls{rmse} for single simulation realization in \cref{appfig:noise_compare2}}
    \label{fig:mse_compare2}
    \vspace{-2em}
\end{figure}

As the targets disperse in free space, the 6 radars coordinate to position 2 radars on either side of each target's direction of motion. This strategy intuitively avoids maintaining a circular formation, which would place the radars too far from the targets, leading to noisy measurements and inaccurate target localization (\cref{appfig:noise_compare2} (a-h)). Optimal radar placement remains relatively similar regardless of whether a distance-independent or distance-dependent range noise model is used, though the radars are positioned farther from the targets in the former case which leads to slightly higher localization error and 90\% \gls{hdi} interval (\cref{appfig:noise_compare2} (c,d,g,h)). In particular, the upper-most radar in \cref{appfig:noise_compare2}(h) is quite far away from the upper right-most target. The extended \gls{mc} simulation over 1000 time steps is shown in \cref{appfig:noise_compare2} with corresponding target localization error depicted in \cref{fig:mse_compare2}.

\subsubsection{4 Mobile Radars and 4 Targets}
We initialize two targets to move towards the upper right and two targets to move towards the lower left. The initial positions of the targets are:
\begin{align*}
    \chi^{T_{1:M}}_0 = \begin{bmatrix}
        0 & 15 & 70 & 15 & 15 & 0 \\
        40.4 & 15 & 70 & 15 & 15 & 0 \\
        -30 & -15 & 45 & -10 & -10 & 0 \\
        20 & -15 & 45 & -10 & -10 & 0 \\
    \end{bmatrix}
\end{align*}
where the first 3 columns are the 3D xyz coordinates, and the last 3 columns are the directional velocities.

We hypothesize that the radars will pair off, with two radars following each group of two targets.

\begin{figure}[h!]
    \vspace{-1em}
    \centering
    \includegraphics[width=0.98\columnwidth]{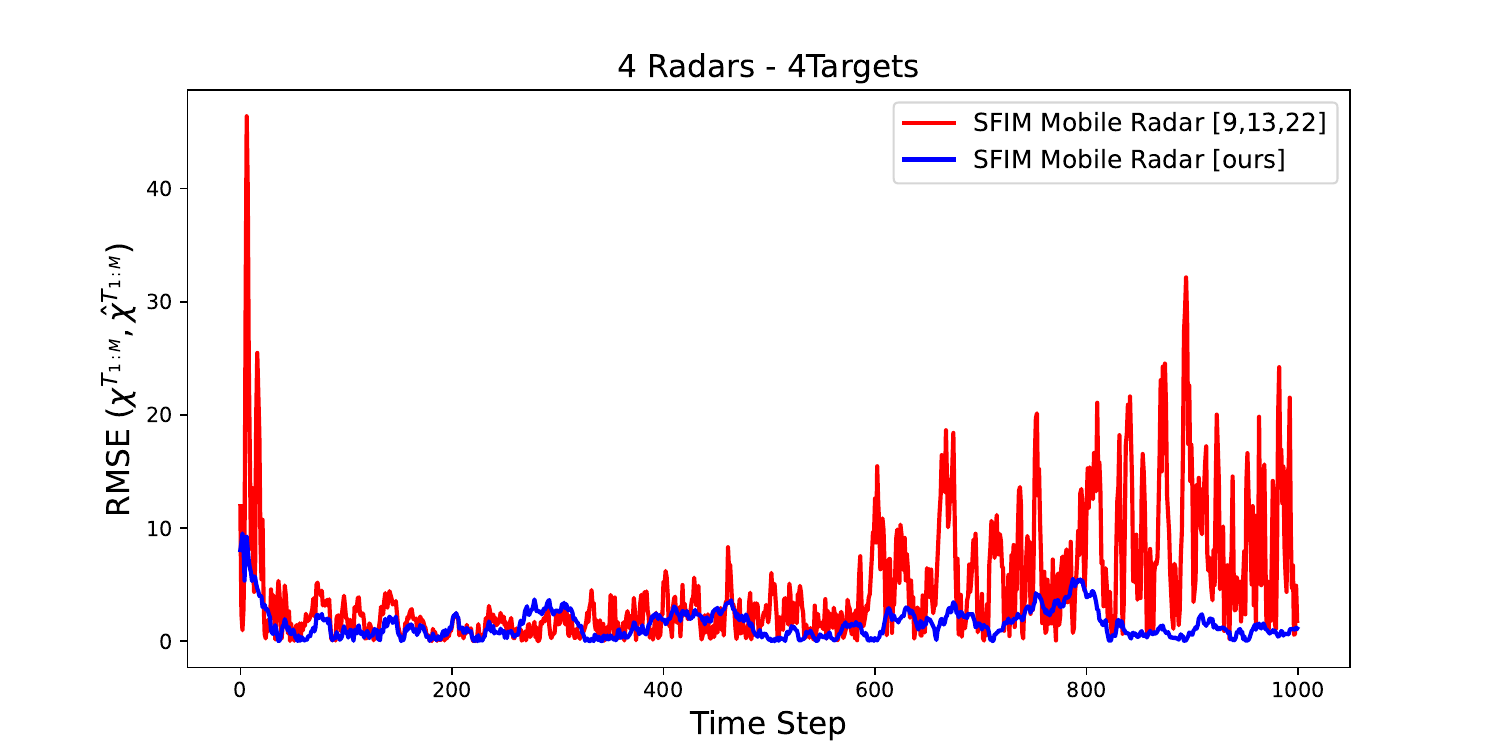}
    \caption{\gls{rmse} for single simulation realization in \cref{appfig:noise_compare3}}
    \label{fig:mse_compare3}
    \vspace{-2em}
\end{figure}

As the targets disperse in free space, the 4 radars coordinate to position 2 radars on either side of each pair of targets' direction of motion. This strategy also avoids maintaining a circular formation, which would place the radars too far from the targets, leading to noisy measurements and inaccurate target localization (\cref{appfig:noise_compare3} (a-h)). Optimal radar placement for the \gls{ccr} measurement model based controller is quite different from the \gls{ddr} measurement model based controller, where the former positions one of the radars extremely far away from all targets in the simulation (\cref{appfig:noise_compare3} (b,d,f,i)) leading to  high localization error. In particular, the upper-left most radar in \cref{appfig:noise_compare3}(h) is quite far all targets. The extended \gls{mc} simulation over 1000 time steps is shown in \cref{appfig:noise_compare3} with corresponding target localization error depicted in \cref{fig:mse_compare3}.

\begin{figure}[htp!]
\centering
\subfigure[3 radars and 4 targets. The uppermost target shows significant height error, while the leftmost target exhibits substantial position error in the xy plane.]{\includegraphics[width=1.0\columnwidth]{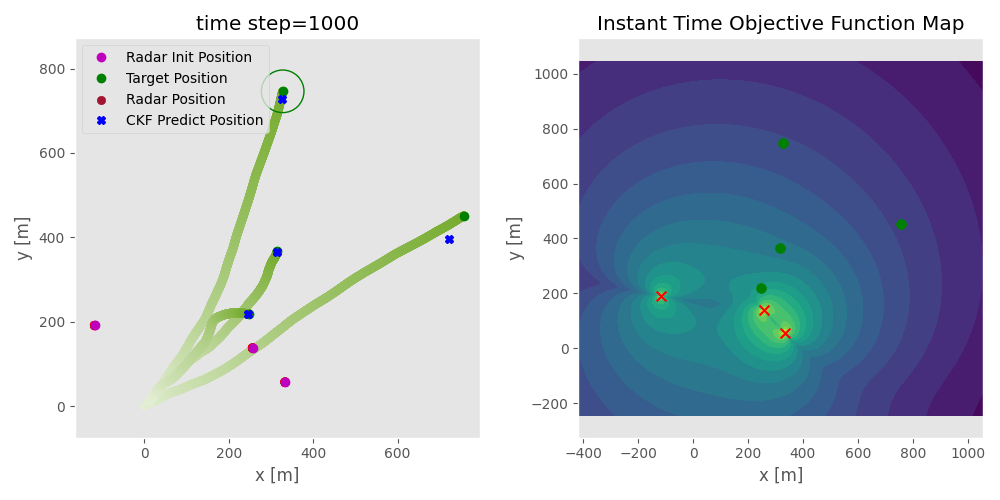}}\hfil  
\subfigure[6 radars and 3 targets. The bottom rightmost target exhibits significant height error, while the top rightmost target displays a position error in the xy plane.]{\includegraphics[width=1.0\columnwidth]{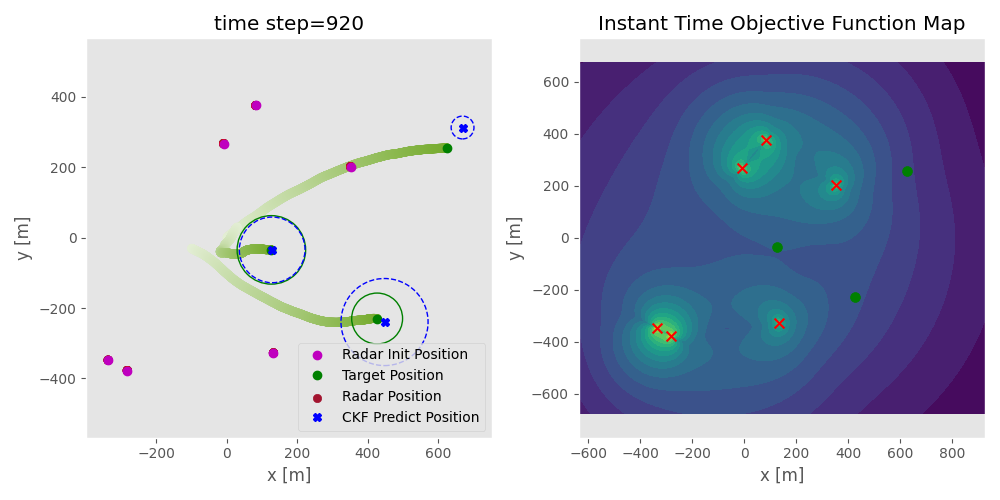}}\hfil  
\subfigure[4 radars and 4 targets. The right uppermost target exhibits significant height error.]{\includegraphics[width=1.0\columnwidth]{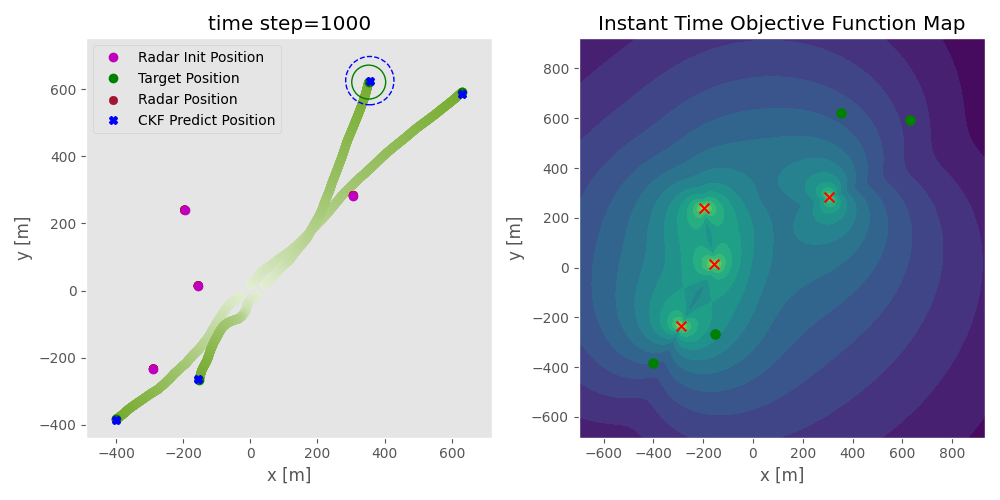}}\hfil  
\vspace{-0.5em}
\caption{Snapshots of each simulation scenario for stationary radars highlight the poor geometric layout of stationary radars for target localization.}
\label{fig:stationary_radars}
\vspace{-2.5em}
\end{figure}

\subsubsection{Stationary Radars}
Mobile radar methods, whether employing constant covariance or distance-dependent covariance in the range measurement model, outperform stationary radars (\cref{fig:rmse_compare} (a-i)). We present a qualitative example for each simulation scenario, illustrating the failure of stationary radar to track targets (\cref{fig:stationary_radars} (a-c)) and showing how radar positioning affects the information gain of measurements. As targets move farther away from the stationary radars, the noise in the range measurements increases drastically, resulting in higher localization errors. 

\glsresetall
% \pagebreak
\vspace{-1em}
\section{CONCLUSION}
We implemented two improvements to optimize radar placement for multi-target localization: a more realistic radar range measurement model and an efficient stochastic optimal controller (\gls{mppi}) for handling non-smooth, non-convex objectives in real time. Our approach significantly improves the \gls{ckf} target state estimate \gls{rmse} compared to stationary radars. Additionally, we show the pitfalls of the \gls{ccr} measurement model, which may increase target localization error by moving radars in the opposite direction of target movement to optimize angles rather than the radar-target distance. Overall, the proposed strategy outperforms stationary radars and simplified range measurement models in target localization, achieving a 38-74\% reduction in mean \gls{rmse} and an 33-79\% reduction in the upper tail of the 90\% \gls{hdi} over 500 \gls{mc} trials across all time steps. For future work, we plan to integrate occupancy grid maps into the collision cost function to account for static and dynamic obstacles, such as trees and buildings, and further improve the radar range measurement model with traditional \gls{mtt} characteristics.

% GS: Acknowledgement goes up front for this journal, 
% I just noticed when disabling other content
%\section*{ACKNOWLEDGMENT}

\bibliography{bibliography}

\begin{IEEEbiography}[{\includegraphics[width=0.6in,height=1.25in,clip,keepaspectratio]{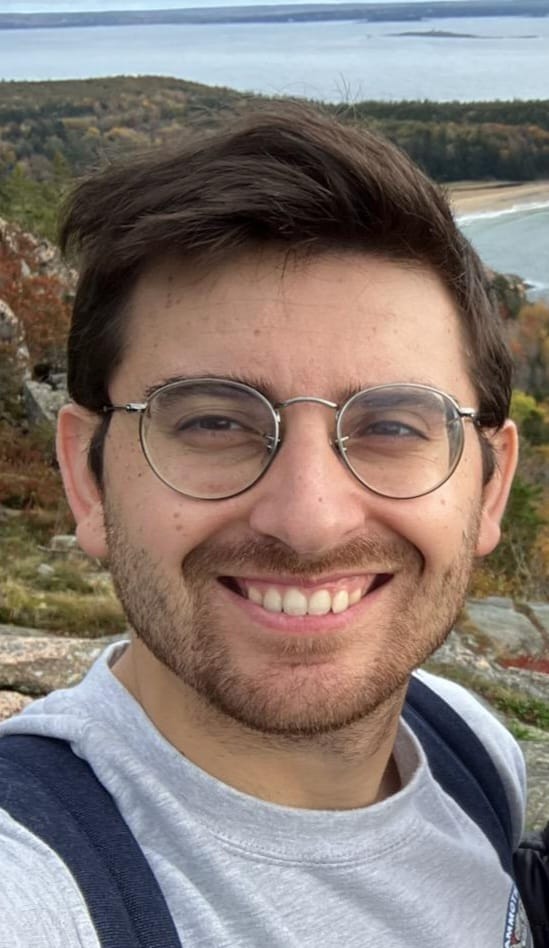}}]{Michael Potter} is a Ph.D. student at Northeastern University (NEU) advised by Deniz Erdo\u{g}mu\c{s} of the Cognitive Systems Laboratory (CSL). He received his B.S, M.S., and M.S.  degrees in Electrical and Computer Engineering from NEU and University of California Los Angeles (UCLA) in 2020, 2020, and 2022 respectively. His research interests are Bayesian Neural Networks, uncertainty quantification, and dynamics based manifold learning. 
\end{IEEEbiography}%
\vspace{-1em}

\begin{IEEEbiography}[{\includegraphics[width=1in,height=1.25in,clip,keepaspectratio]{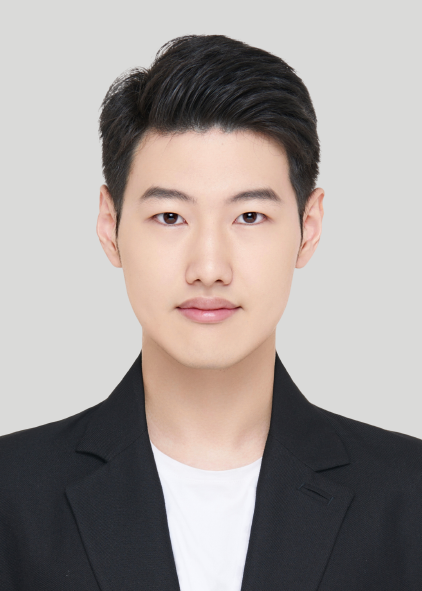}}]{Shuo Tang}{\space} received the BS degree in mechanical engineering from China Agricultural University, China and the MS degree in mechanical engineering from Northeastern University, Boston, MA, in 2018 and 2020, respectively.
He is currently a PhD candidate in electrical and computer engineering at Northeastern University. His research interests include GNSS signal processing, sensor fusion and computational statistics.
\end{IEEEbiography}%
\vspace{-1em}

\begin{IEEEbiography}[{\includegraphics[width=1in,height=1.25in,clip,keepaspectratio]{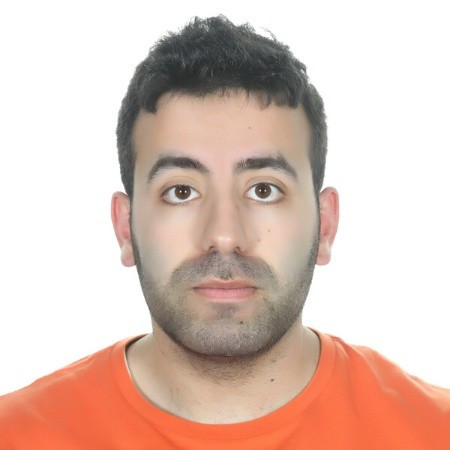}}]{Paul Ghanem}{\space} received the B.E. and M.S. degrees in Electrical Engineering and Robotics in 2016 from the Lebanese American University and in 2019 from the University of Maryland, College Park. He is currently a PhD student at Northeastern University. He is a research assistant at the cognitive systems Lab at NEU. His research interests are in the area of machine learning, robotics and automatic Control.
\end{IEEEbiography}%
\vspace{-1em}

\begin{IEEEbiography}[{\includegraphics[width=1in,height=1.25in,clip,keepaspectratio]{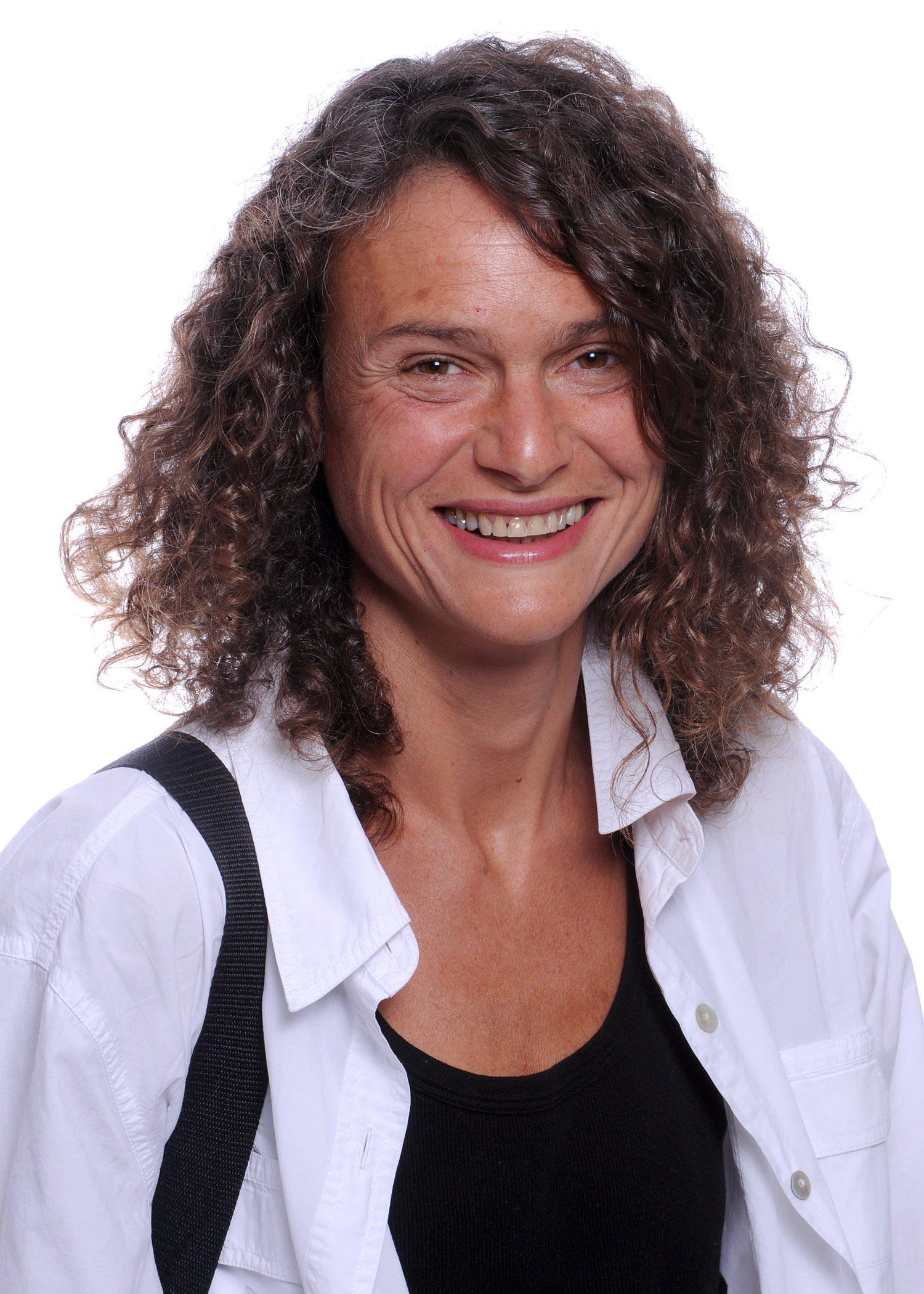}}]{Milica Stojanovic}(Fellow, IEEE), graduated from the University of Belgrade, Serbia, in 1988, and received M.S. (’91) and Ph.D. (’93) degrees in electrical engineering from Northeastern University, Boston, Massachusetts. She was a Principal Scientist at the Massachusetts Institute of Technology, and in 2008 joined Northeastern University, where she is currently a Professor of electrical and computer engineering. She is also a Guest Investigator at the Woods Hole Oceanographic Institution. 
%Her research interests include digital communications theory, statistical signal processing and wireless networks, and their applications to underwater acoustic systems. 
She is an Associate Editor for the IEEE Journal of Oceanic Engineering, a past Associate Editor for the IEEE Transactions on Signal Processing, IEEE Transactions on Vehicular Technology, and a Senior Editorial Board Member of the IEEE Signal Processing Magazine.  She chairs the IEEE Ocean Engineering Society’s (OES) Technical Committee for Underwater Communication, Navigation and Positioning, and is an IEEE OES Distinguished Lecturer. Milica is the recipient of the 2015 IEEE/OES Distinguished Technical Achievement Award, the 2019 IEEE WICE Outstanding Achievement Award, and the 2023 IEEE Communications Society’s Women Stars in Computer Networking and Communications Award. In 2022, she was awarded an honorary doctorate from the Aarhus University in Denmark, and was elected to the Academy of Engineering Sciences of Serbia. 
\end{IEEEbiography}
\vspace{-1em}
\begin{IEEEbiography}[{\includegraphics[width=1in,height=1.25in,clip,keepaspectratio]{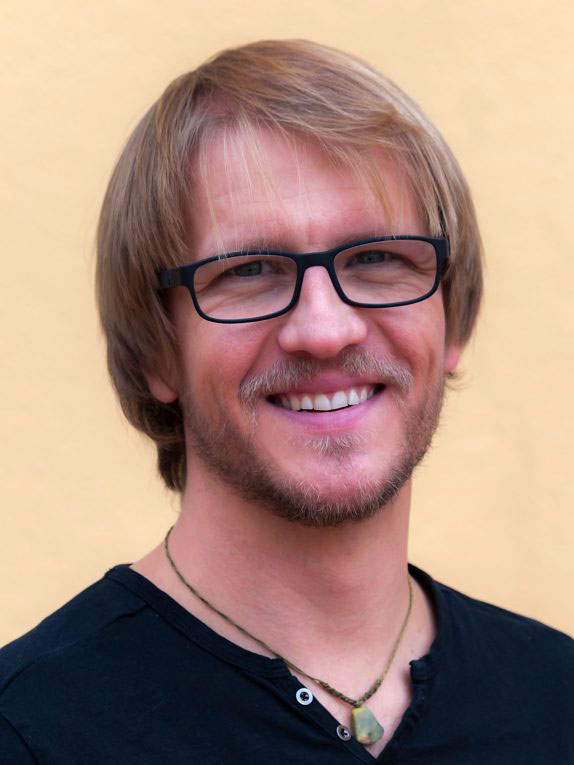}}]{Pau Closas}(Senior Member, IEEE),
is an Associate Professor in Electrical and Computer Engineering at Northeastern University, Boston MA.
He received the MS and PhD in Electrical Engineering from UPC in 2003 and 2009, respectively. He also holds a MS in Advanced Maths and Mathematical Engineering from UPC since 2014. He is the recipient of the EURASIP Best PhD Thesis Award 2014, the $9^{th}$ Duran Farell Award for Technology Research, the $2016$ ION's Early Achievements Award, $2019$ NSF CAREER Award, and the IEEE AESS Harry Rowe Mimno Award in $2022$. His primary areas of interest include statistical signal processing, stochastic filtering, robust filtering, and machine learning, with applications to positioning and localization systems. He is the EiC of the IEEE AESS Magazine, volunteered in other editorial roles (e.g. NAVIGATION, Proc. IEEE, IEEE Trans. Veh. Tech., and IEEE Sig. Process. Mag.), and has been actively involved in organizing committees of a number of conference such as EUSIPCO (2011, 2019, 2021, 2022), IEEE SSP'16, IEEE/ION PLANS (2020, 2023), or IEEE ICASSP'20.
\end{IEEEbiography}

\begin{IEEEbiography}[{\includegraphics[width=1in,height=1.25in,clip,keepaspectratio]{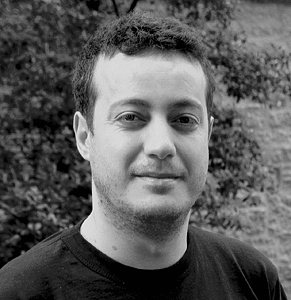}}]%
{Murat Akcakaya} (Senior Member, IEEE)
received his Ph.D. degree in Electrical Engineering from the Washington University in St. Louis, MO, USA, in December 2010. He is currently an Associate Professor in the Electrical and Computer Engineering Department of the University of Pittsburgh. His research interests are in the areas of statistical signal processing and machine learning.
\end{IEEEbiography}
\vspace{-1em}

\begin{IEEEbiography}[{\includegraphics[width=1in,height=1.25in,clip,keepaspectratio]{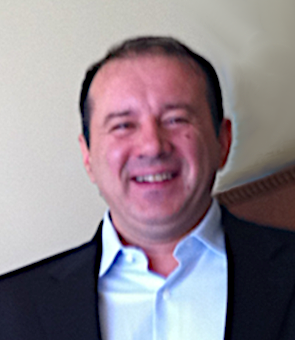}}]
{Marius Necsoiu} (Member, IEEE) {\space}received his PhD in Environmental Science (Remote Sensing) from the University of North Texas (UNT), in 2000. He has broad experience and expertise in remote sensing systems, radar data modeling, and analysis to characterize electromagnetic environment behavior and geophysical deformation. As part of the DEVCOM ARL he leads research in cognitive radars and explores new paradigms in AI/ML science that are applicable in radar/EW research.
\end{IEEEbiography}
\vspace{-1em}

\begin{IEEEbiography}
[{\includegraphics[width=1in,height=1.25in,clip,keepaspectratio]{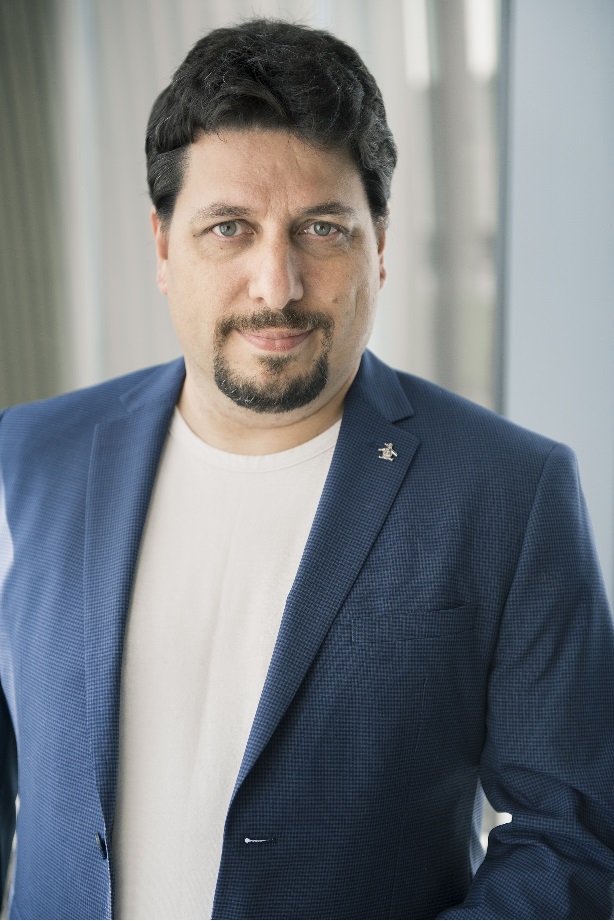}}]
{Deniz Erdo\u{g}mu\c{s}} (Sr Member, IEEE), received BS in EE and Mathematics (1997), and MS in EE (1999) from the Middle East Technical University, PhD in ECE (2002) from the University of Florida, where he was a postdoc until 2004. He was with CSEE and BME Departments at OHSU (2004-2008). Since 2008, he has been with the ECE Department at Northeastern University. His research focuses on statistical signal processing and machine learning with applications data analysis, human-cyber-physical systems, sensor fusion and intent inference for autonomy. He has served as associate editor and technical committee member for multiple IEEE societies.
\end{IEEEbiography}

\vspace{-1em}

\begin{IEEEbiography}
[{\includegraphics[width=1in,height=1.25in,clip,keepaspectratio]{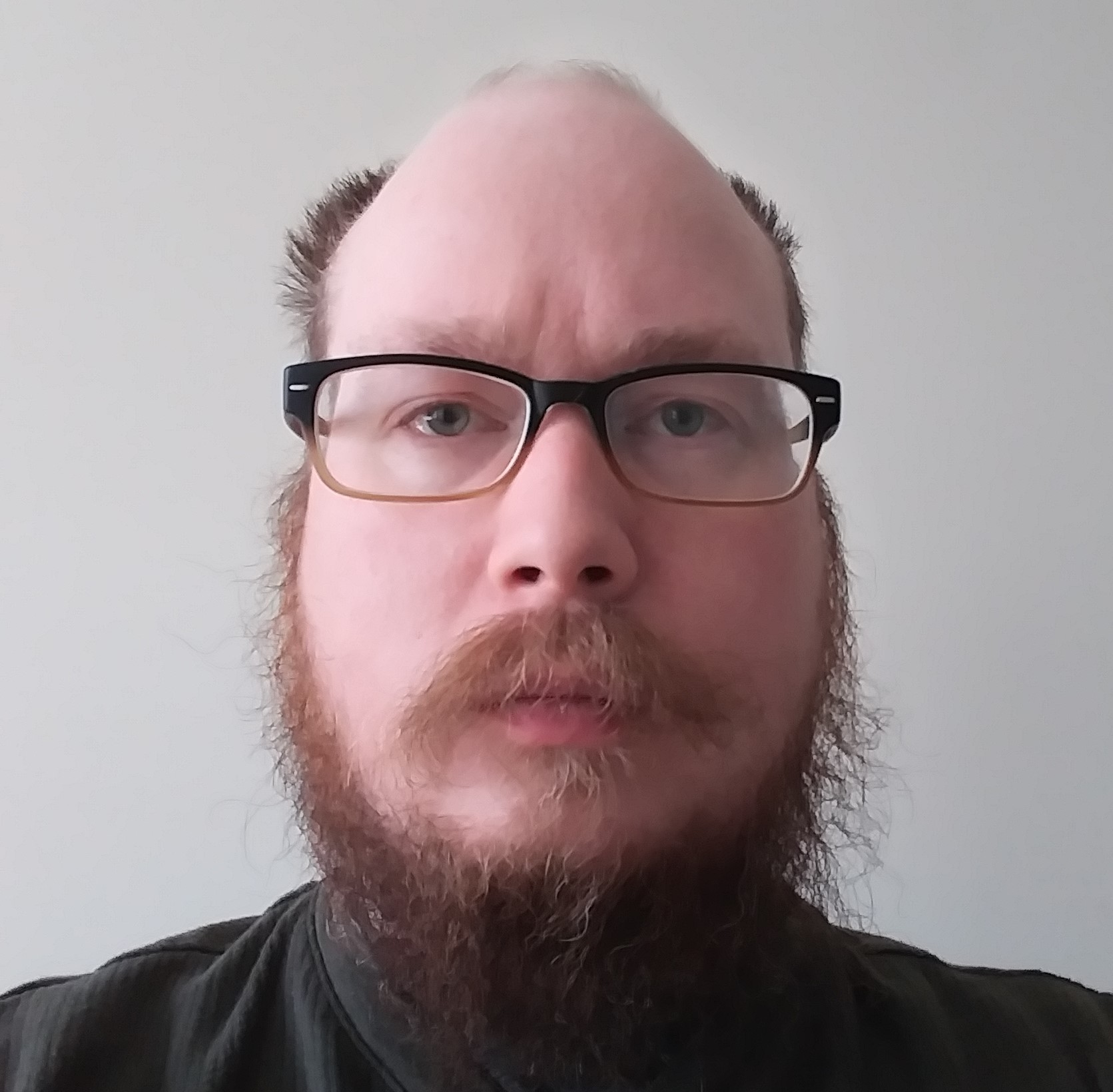}}]
{Ben Wright} is a Research Engineer at Kostas Research Institute. He received his Ph.D. in Computer Science from New Mexico State University in 2018. He was a Postdoctoral Researcher in the Navy Center for Applied Research in AI at the US Naval Research Laboratory from 2019-2021. His research interests involve Multi-agent Reasoning, Computational Logic, and Human-Robot Interaction.    
\end{IEEEbiography}
\vspace{-1em}

\begin{IEEEbiography}[{\includegraphics[width=1.0in,height=1.25in,clip,keepaspectratio]{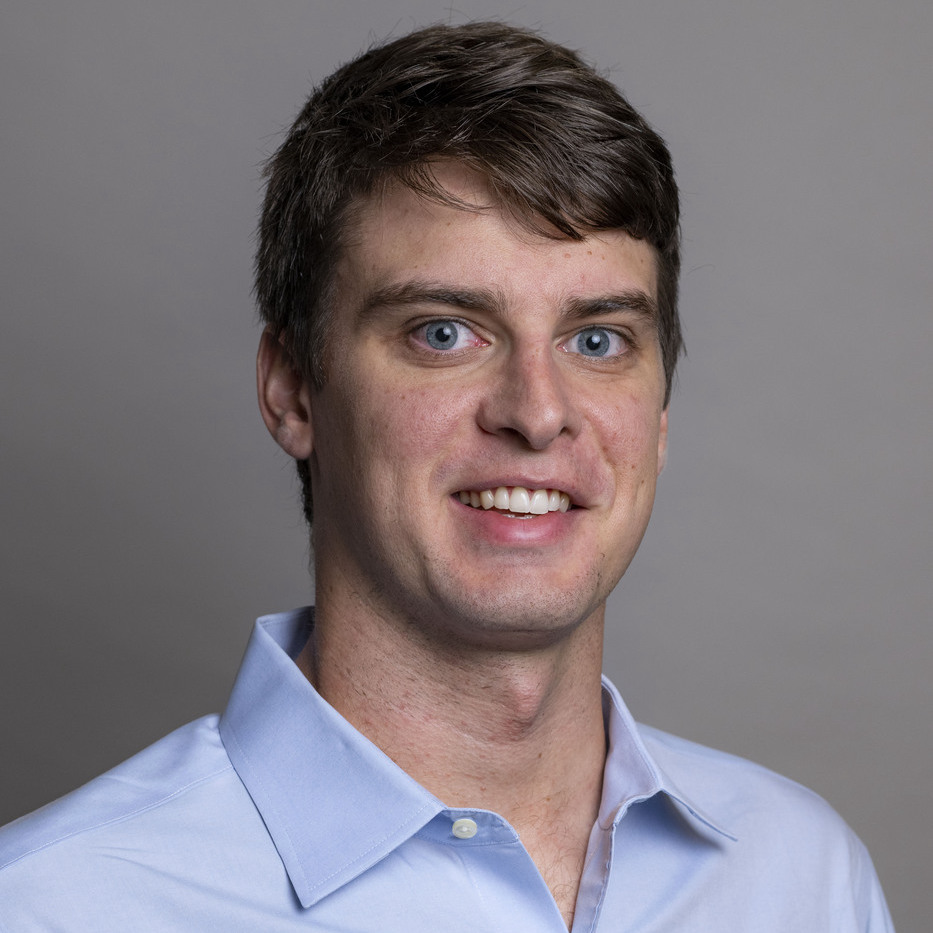}}]{Michael Everett} received the S.B., S.M., and Ph.D. degrees in mechanical engineering from the Massachusetts Institute of Technology (MIT), Cambridge, MA, USA, in 2015, 2017, and 2020, respectively. He was a Post-Doctoral Associate and Research Scientist in the Department of Aeronautics and Astronautics at MIT. He was a Visiting Faculty Researcher at Google Research.
He joined Northeastern University in 2023, where he is currently an Assistant Professor in the Department of Electrical \& Computer 
Engineering and Khoury College of Computer Sciences at Northeastern University, Boston, MA, USA. His research lies at the intersection of machine learning, robotics, and control theory, with specific interests in the theory and application of safe and robust neural feedback loops.
Dr. Everett's work has been recognized with numerous awards, including the Best Paper Award in Cognitive Robotics at IEEE/RSJ International Conference on Intelligent Robots and Systems (IROS) 2019.
\end{IEEEbiography}

\vspace{-1em}

\begin{IEEEbiography}
[{\includegraphics[width=1in,height=1.25in,clip,keepaspectratio]{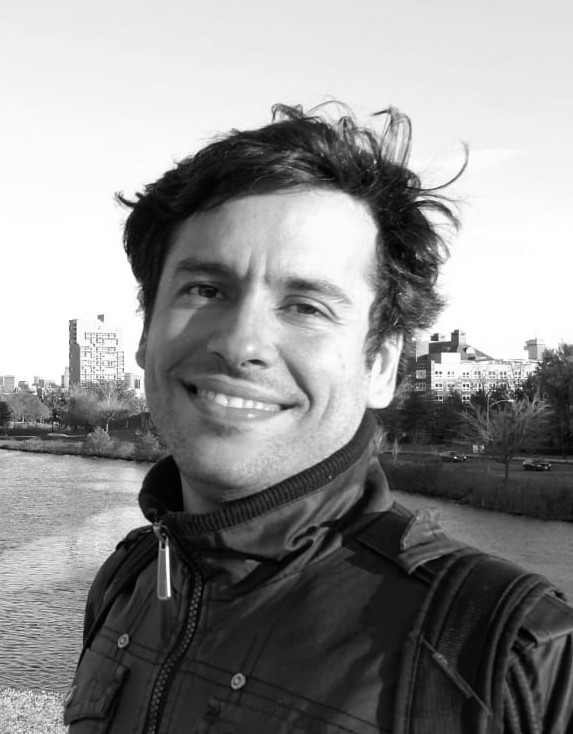}}]%{figures/carnival_floripa_tales}}]
{Tales Imbiriba} (Member, IEEE)   
is an Assistant Professor in Computer Science at University of Massachusetts Boston (UMB). Before joining UMB, he served as an Assistant Research Professor at the ECE dept. and Senior Research Scientist at the Institute for Experiential AI, both at Northeastern University (NU), Boston, MA, USA. He received his Doctorate degree from the Department of Electrical Engineering (DEE) of the Federal University of Santa Catarina (UFSC), Florian\'opolis, Brazil, in 2016. He served as a Postdoctoral Researcher at the DEE--UFSC (2017--2019) and at the ECE dept. of the NU (2019--2021). 
His research interests include audio and image processing, pattern recognition, Bayesian inference, online learning, and physics-guided machine learning.
% \end{IEEEbiographynophoto}
\end{IEEEbiography}

\clearpage
\input{appendix_paper_r1}

\end{document}

%% file: acronyms.tex
\newacronym{mppi}{MPPI}{Model Predictive Path Integral}
\newacronym{mpc}{MPC}{Model Predictive Controller}
\newacronym{fim}{FIM}{Fisher Information Matrix}
\newacronym{pfim}{PFIM}{Posterior Fisher Information Matrix}
\newacronym{sfim}{SFIM}{Standard Fisher Information Matrix}
\newacronym{pcrlb}{PCRLB}{Posterior Cramer Rao Lower Bound}
\newacronym{ckf}{CKF}{Cubature Kalman Filter}
\newacronym{sar}{SAR}{Synthetic Aperture Radar}
\newacronym{soc}{SOC}{Stochastic Optial Control}
\newacronym{ais}{AIS}{Adaptive Importance Sampling}
\newacronym{awgn}{AWGN}{Additive White Gaussian Noise}
\newacronym{ew}{EMW}{Electromagnetic Wave}
\newacronym{uav}{UAV}{Unmanned Aerial Vehicle}
\newacronym{li}{LIDAR}{Light Detection and Ranging}

\newacronym{uhf}{UHF}{Ultra High Frequency}
\newacronym{sota}{SOTA}{state-of-the-art}
\newacronym{mtt}{MTT}{Multi-Target Tracking}
\newacronym{jpdaf}{JPDAF}{Joint Probabilistic Data Association Filter}

\newacronym{pdf}{PDF}{Probability Density Function}
\newacronym{blue}{BLUE}{Best Linear Unbiased Estimator}

\newacronym{pf}{PF}{Particle Filter}
\newacronym{kbl}{KL-Divergence}{Kullback-Leibler Divergence}
\newacronym{oas}{OAS}{Oracle Approximating Shrinkage}
\newacronym{rmse}{RMSE}{Root Mean Squared Error}

\newacronym{mc}{MC}{Monte Carlo}
\newacronym{snr}{SNR}{Signal Noise Ratio}
\newacronym{ofdm}{OFDM}{Orthogonal Frequency Division Multiplexing}
\newacronym{psk}{PSK}{Phase Shift Keying}
\newacronym{lw}{LW}{Ledoit-Wolf}
\newacronym{ccr}{CCR}{constant covariance range}
\newacronym{ddr}{DDR}{distance dependent range}
\newacronym{hdi}{HDI}{Highest Density Interval}
\newacronym{ecdf}{ECDF}{Emperical Cumulative Distribution Function}

%% file: introduction/introduction_draft_r1.tex
\section{INTRODUCTION}
\label{sec:intro}

\begin{figure*}[h!]
    \centering
    \subfigure[Search and rescue in forest fire]{\includegraphics[width=5cm]{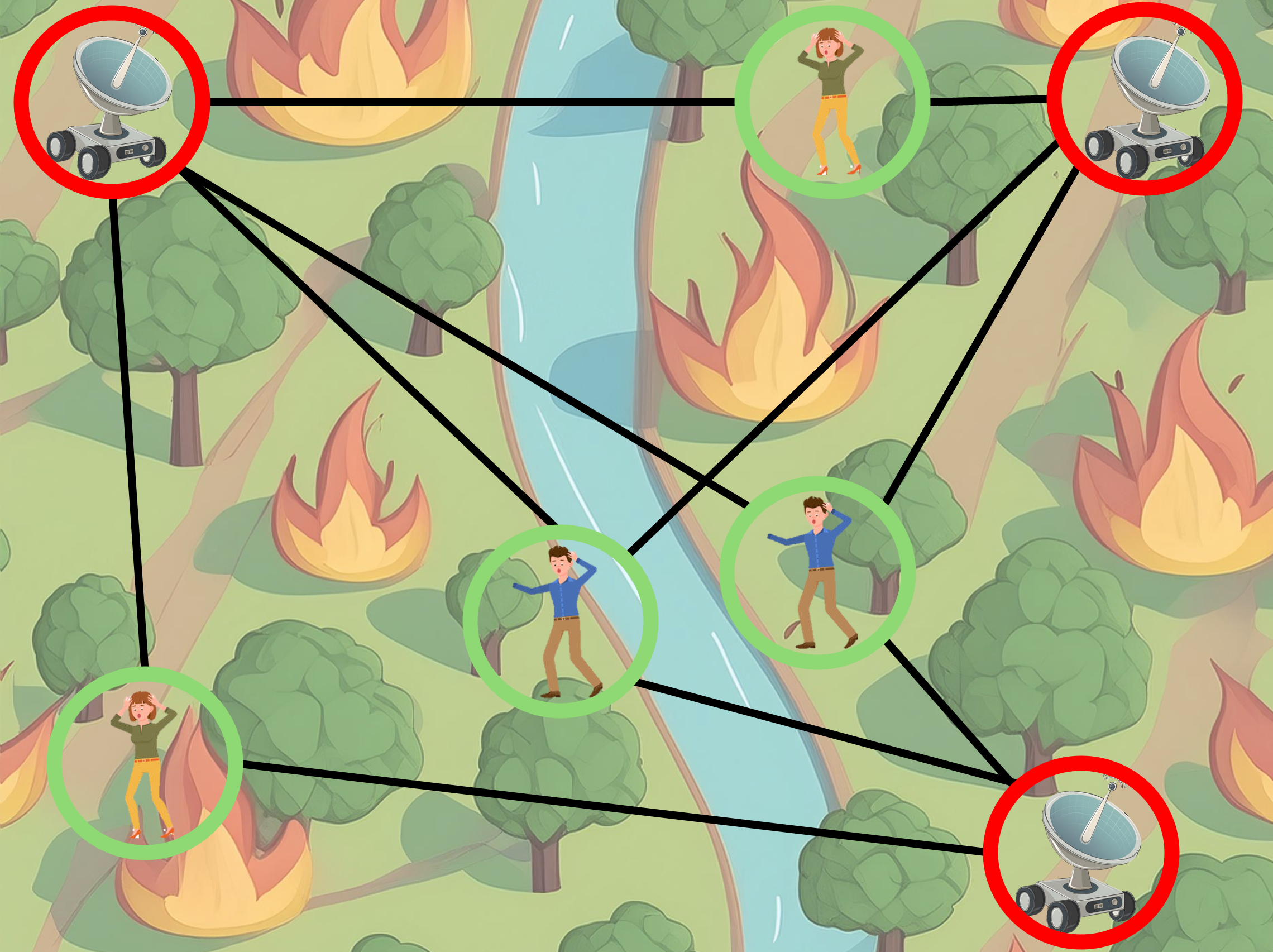}} \hfil 
    \subfigure[Enemy drone localization and tracking]{\includegraphics[width=5cm]{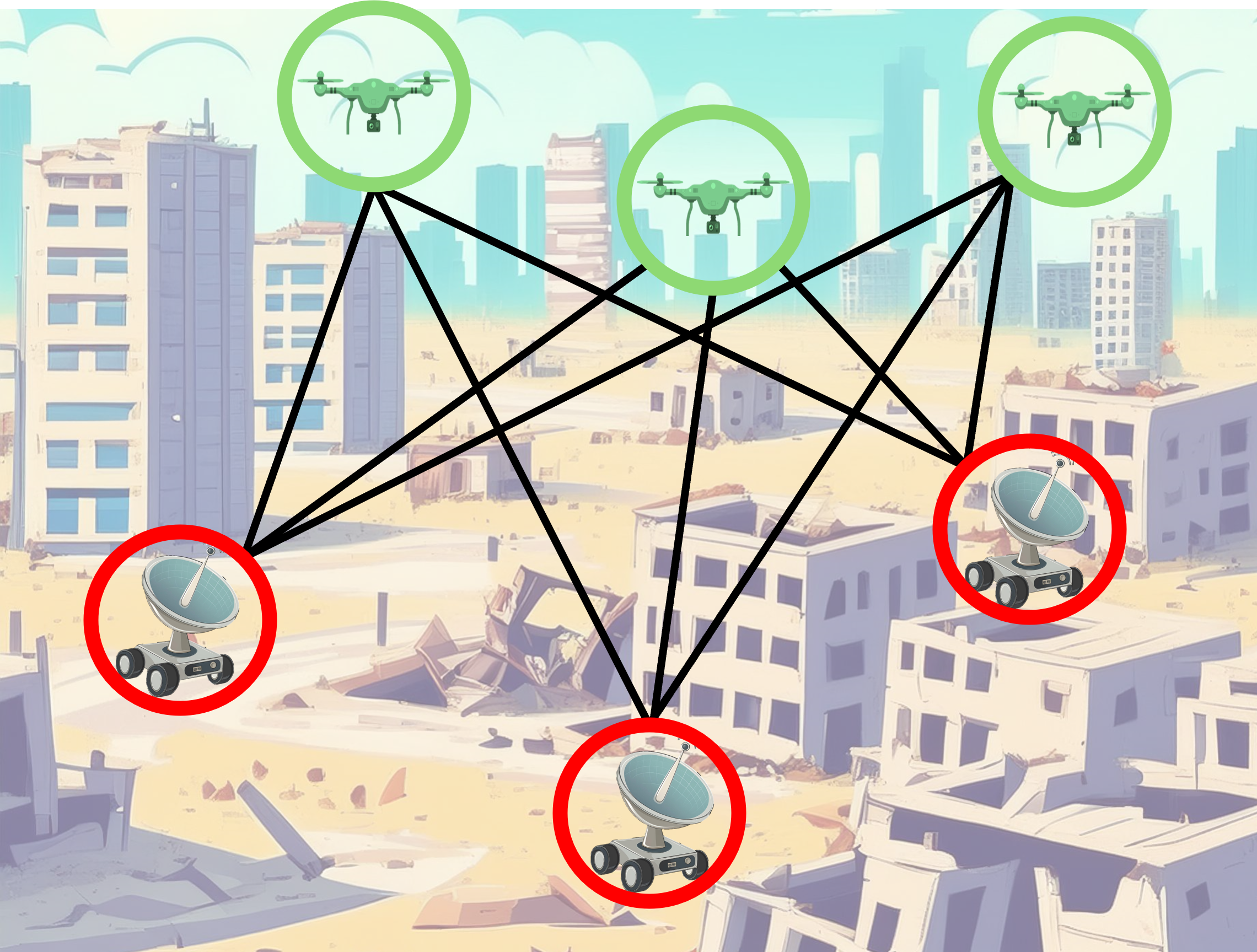}}\hfil  
    \caption{Subfigure (a) shows the civilian application of a search and rescue mission where mobile radars navigate through a dense forest (on fire) to quickly localize and track lost victims. Subfigure (b) shows the military application of enemy drone localization and tracking in a city environment during  a cloudy day.}
    \label{fig:conops}
    \vspace{-1.5em}
\end{figure*}

Deploying a fleet of autonomous mobile sensors to estimate the positions and velocities of targets over time is an important research problem, with applications in transportation, navigation, defense, surveillance, and emergency rescue~\cite{ding2023radar,venkatasubramanian2023datadriven,grant2010radar,kumar2021recent}, exemplified in \cref{fig:conops}. This problem involves multiple challenges, including predicting expected sensor information gain without exact target or measurement data, adapting mobile sensors planned trajectories to dynamic environmental and target changes, and selecting suitable sensor models and modalities. We will discuss these challenges in the following paragraphs.

The multi-target tracking and sensor placement literature has explored various sensing modalities, including cameras \cite{yoon2009autonomous}, ultrasonic sensors \cite{juan2021object,ray2002genetic}, \gls{li} \cite{9810346}, acoustic range sensors \cite{crasta2018multiple}, and radar-based techniques dating back to World War II.
This paper focuses on tracking with radar in the \gls{uhf} (particularly ``cognitive radars'' \cite{gurbuz2019overview}), due to the long-range detection capability, ability to penetrate atmospheric obstacles, and versatility in capturing detailed target features \cite{richards2010principles}.
% Our approach consists of algorithm estimating target states from noisy radar data, while simultaneously controlling the position of the radars to improve the target estimates without

% This target localization problem 
% has long relied on radars in military operations, dating back to World War II, where radar positioning was crucial for mission success \cite{butowsky1994opana}. Ideally, radars would dynamically adjust positions to optimize coverage of enemy combatants while minimizing detection risks, thereby increasing the likelihood of mission success. That being said, the accuracy of target localization hinges on factors like sensing technology and sensor-target geometry, highlighting the importance of understanding and optimizing these relationships \cite{godrich2010target,sadeghi2021target,sensingtech}.
% Radars are ubiquitous in target detection and localization 

% need to relate radar/target geometry to target localization performance -> use PCRLB/PFIM -> most people only do this with overly simple sensor model

A key challenge in radar-based target tracking is in incorporating realistic sensor models into the information metrics used for controlling the sensors.
In particular, most existing control strategies for this problem use the \gls{pfim} \cite{tichavsky1998posterior} (inverse of the \gls{pcrlb}) to quantify the relationship between radar-target geometry and target localization performance \cite{bishop2010optimality,moreno2011optimal,zhao2002information}. However, studies optimizing sensor placement often oversimplify by using a constant covariance matrix for range measurements, neglecting signal path attenuation effects \cite{kumar2021recent}.  The approach in \cite{kumar2021recent} focuses on optimizing the angle subtended between the radars and targets but does not encourage the mobile radars to move closer to the target. In \cite{godrich2010target}, the authors provide the error covariance matrix of a \gls{blue} estimator for the range measurement using a first-order Taylor Series approximation of a \gls{ddr} measurement model, which yields the \gls{fim}. Similarly, we calculate the \gls{fim} for the \gls{ddr} measurement model from \cite{godrich2010target}, where the covariance matrix of the range measurement is directly proportional to the received power at the radar after signal path attenuation, but we do not apply a first-order Taylor Series approximation to the \gls{ddr} measurement model. This strategy encourages mobile radars to move closer to the targets, thereby increasing the \gls{snr} and improving radar range measurements.

% dynamic environment -> need mpc -> hard to solve because of nonlinear FIM objective -> propose to use mppi
Another major challenge in scenarios with moving targets and movable radars is that the optimal radar-target geometry may change substantially over time.
Consequently, many studies use \gls{mpc} \cite{mpc_matlab} for choosing the control inputs of mobile sensors over a finite horizon in response to uncertain target state estimates \cite{khan2022design,uluskan2020noncausal,chu2020trajectory,sture2020trajectory,hung2020range}. However, natural choices for the \gls{mpc} objective in this problem, such as minimizing/maximizing the  D-optimality of \gls{pcrlb}/\gls{pfim}, lead to challenging nonlinear optimization problems that are computationally prohibitive to solve exactly.
To address this issue, we instead utilize \gls{mppi} to approximate the optimal control input distribution for the mobile radars. \gls{mppi} enables the handling of discontinuous, non-smooth objectives through highly parallelized computation for real-time control \cite{mppi_advantages,williams2018information}.
% Typically, the number of samples needed to estimate the parameters of the \gls{pdf} with a desired level of accuracy grows exponentially with the dimensionality of the parameters, known as the curse of dimensionality \cite{bishop2006pattern}.
However, as the number of agents, planning horizon, and the actuation space increases, parameter estimation, particularly for the covariance, requires an exponentially growing number of samples. To mitigate this issue and reduce the number of proposal distribution updates and trajectory samples needed in the high-dimensional control space, we integrate advanced \gls{ais} techniques, such as cross-entropy weighting with covariance shrinkage \cite{asmar2023model}, to prevent overfitting of the sample covariance matrix.

% optimizing \gls{mpc} objectives for nonlinear functions, particularly those involving \gls{pcrlb}/\gls{pfim}, may be computationally intensive and impose restrictive assumptions. Thus, 

This paper introduces a improved pipeline for continuously optimizing radar placement, aiming to overcome the previously mentioned challenges. We focus on range-only radars, deriving objective functions using \glspl{fim} that incorporate signal path attenuation in the range measurement model. These functions are optimized with respect to radar control inputs using \gls{mppi} control. Crucially, \gls{mppi} control, which is rapidly gaining popularity in control research, offers enhanced flexibility for real-time optimization of nonlinear, discontinuous, and non-differentiable objectives. These advantages enable \gls{mppi} to handle future extensions of the proposed objective, such as incorporating more complex signal measurement models or adding more constraints through occupancy grid maps. 
%Results from over 500 \gls{mc} trials across various simulation scenarios empirically show that our approach achieves lower \gls{rmse} compared to stationary radars and state estimators using simplified range measurement models, with reduced variance across \gls{mc} trials. 

The paper is structured as follows: \cref{sec:related_works} describes the current literature; \cref{sec:preliminaries} describes the simulation assumptions and notation; \cref{sec:radar_signal_model} describes the radar signal model and \gls{ddr} measurement model; \cref{sec:process_modeling} formulates the targets' state-space model and the radars' kinematic model; \cref{sec:fim_definition} derives the \gls{fim} which accounts for the radar signal attenuation; \cref{sec:mpc_objective} formulates our objective function based on the derived \glspl{fim} and outlines our \gls{mppi} controller, and \cref{sec:experiment} discusses the results of the proposed approach for three simulation scenarios.

%% file: relatedwork/relatedwork_draft_r1.tex
\section{Related Work}
\label{sec:related_works}

We explore the literature on optimizing sensor trajectories using the \gls{fim} of the likelihood function of the range measurements, where the likelihood function parameters correspond to the target positions. Various forms of the \gls{fim} establish a fundamental estimation lower bound limit for the mean squared error covariance matrix \cite{bishop2006pattern}. Therefore, many authors construct an objective as the D/A-optimality \cite{jones2021optimal} of the \gls{pfim}, parameterized by sensor and estimated target positions \cite{wireless_network_pos}. We divide these works into two categories: those that consider static-sensors versus those that consider dynamic-sensors.

\vspace{-1em}
\subsection{Static Sensors}
Static sensors lack dynamic models or control inputs. Optimizing their placement involves determining where they would be positioned if placed instantaneously. Many studies on optimal radar placement use range measurements assumed to follow Gaussian distributions to construct the \gls{fim}, where the covariance may or may not vary with sensor-target distance \cite{jiang2015optimal,bishop2010optimality,moreno2011optimal,godrich2010target,hu2017cramer,yan2017exact}.

Several studies (e.g., \cite{jiang2015optimal,bishop2010optimality}) suggest that when the range measurement covariance is distance-independent, optimal sensor placement depends only on the angle between the sensor and the target. However, this approach overlooks the impact of signal path attenuation, which ideally results in higher information gain when the sensor is closer to the target.

To address this, \cite{moreno2011optimal,jourdan2008optimal} adopt a heuristic assumption concerning covariance, combining a constant diagonal matrix with another diagonal matrix which is a function of the range between sensors and targets. However, signal processing fundamentals provide more realistic range measurement distributions. We take a similar approach to \cite{godrich2010target,yan2017exact}, where the covariance, which scales with the return radar power, is determined by optimizing the Taylor series expansion of a match filter over time delay for a Narrowband return signal subject to attenuation and passing through an \gls{awgn} channel.
\vspace{-1em}
\subsection{Dynamic Sensors}
Practically, radars cannot simply ``appear'' at the optimal location; they require constrained dynamics and path planning. Moreover, dynamic sensors are essential to track dynamic targets, as the optimal radar configuration at one time step may not remain optimal at the next time step.

Historically, two main approaches have been used for optimal radar placement in target localization of dynamic targets: 1) moving the radars to track dynamic targets while maintaining an optimal radar-target localization geometry \cite{martinez2006optimal,kumar2021recent}, and 2) ``spraying'' radars over a large area to ensure high spatial coverage and multiplicity, then selecting which radars to employ for sensing \cite{zhao2002information}. This paper primarily focuses on the first approach, which is more appropriate given a limited number of radars.

Since the objective is a function of a finite horizon of time steps, many studies utilize an \gls{mpc} \cite{mpc_matlab} objective for optimizing the control inputs of mobile sensors \cite{khan2022design,uluskan2020noncausal,chu2020trajectory,sture2020trajectory,hung2020range}. Consequently, rather than directly optimizing the radar's position at a specific time step, as in the static-sensor formulation, the optimal radar trajectory is indirectly determined by optimizing the control inputs to the radar kinematic model using the method of direct shooting \cite{directshooting}.

% integrating the \gls{pcrlb}/\gls{pfim} in the instantaneous cost function.

The finite horizon \gls{mpc} objective, formulated in terms the \gls{pfim}, may be efficiently computed through recursive evaluation using discrete-time nonlinear filtering \cite{tichavsky1998posterior}. Several studies on cooperative range-based underwater target localization with multiple autonomous surface vehicles adopt this approach, while also considering constraints such as distances between targets and sensors \cite{crasta2018multiple,hung2020range,sture2020trajectory}. However, these methods often assume a constant diagonal covariance matrix for range measurements, irrespective of range, typically involve slow-moving sensors and targets, assume the target remains at a constant depth, and may assume a zero covariance for the target dynamic model.
Moreover, solving the resulting nonlinear optimization problems remains challenging, with several recent approaches leveraging \texttt{CasADi}~\cite{andersson2019casadi}.
% Moreover, these papers rely on the nonlinear optimization package \cite{andersson2019casadi} for solving the \gls{mpc}, which could impose limitations on the objective formulation and hinder real-time capabilities.

He et al \cite{he2020trajectory} employs a more complex measurement model which integrates \gls{mtt} characteristics such as data association, track insertion, and track deletion using the \gls{jpdaf}. However, this work assumes a constant diagonal covariance matrix for range and bearing measurements. Furthermore, the trajectory optimization only optimizes the sensor's heading angle one time step ahead due to the highly nonlinear objective function, potentially leading to myopic and suboptimal trajectories.

Similar to information-based \gls{mpc} objectives like the D/A optimality of the \glspl{fim}, \cite{cai2021non} minimizes the mutual information gain between the target state distributions before and after the update step of a linear Kalman Filter, while parameterizing the measurement noise covariance based on the robot’s state. However, the measurement covariance only scales linearly with the sensor-target distance and is clipped at an extremely small variance of 0.01 meters, restricting experiments to small square arenas of 40 to 60 meters.

Hence, the proposed approach aims to enhance target localization through continuously optimizing radar placement by integrating an improved range measurement model into the objective function. Additionally, we leverage \gls{mppi} control to increase flexibility in solving the \gls{mpc} objective over non-smooth and highly nonlinear objectives, enabling real-time control capabilities.

\vspace{-2em}

%% file: methodology/methodology_draft_r1.tex
\section{Radar Signal Model}
\label{sec:radar_signal_model}
Following the approach in \cite{godrich2010target}, we assume there are $N$ radars transmitting orthogonal narrowband signals to track $M$ targets. In this setup, the received signal at radar $n$ for target $m$ is given by:
\begin{align}
    r^{nm}(t) &= \sum_{b=1}^N \zeta^{nm} e^{-j2\pi f_c \tau^{nm}} s^b(t-\tau^{nm}) + \epsilon_{a}^n(t), \label{eqn:return_signal} 
\end{align}
where the time delay is:
\begin{align}
    \tau^{nm} &= \frac{\lVert \chib^{R_n}_{xyz} - \chib^{T_m}_{xyz} \rVert  +  \lVert \chib^{R_b}_{xyz} - \chib^{T_m}_{xyz} \rVert} {c},
\end{align}
and the set of orthonormal lowpass equivalent signals $s(t)$ such that there is no signal interference when multiple radars transmit simultaneously.
The return signal noise $\epsilon_{a}(t)$ is circularly symmetric, zero-mean, complex Gaussian noise with autocorrelation function $\sigma_a^2 \delta (\tau)$. $\zeta$ is the amplitude of the return signal, where $|\zeta|^2$ is the radar power return specified by the radar equation \cite{barton2013radar}:
\begin{align}
    |\zeta|^2 = P_r = \frac{P_t \Lambda_t \Lambda_r \lambda^2 \Xi}{(4\pi)^3 \lVert \chib^{R}_{xyz} - \chib^{T}_{xyz} \rVert ^4 L}
\end{align}
The radar parameters specifying the received power $P_r$ $[W]$ at the radar include $P_t$ as the transmit power $[W]$, $\Lambda_t$ and $\Lambda_r$ as the transmit and receive gains, $L$ as the general loss factor, $\Xi$ as the radar cross section $[m^2]$, and $\lambda$ as the carrier frequency wavelength $[m]$.

It has been shown that the range measurement between radar $n$ and target $m$ approximately follows a Gaussian \gls{pdf} with variance parameterized by the radar-target geometry \cite{bishop2010optimality}.
\begin{align}
z^{nm} \sim  \mathcal{N}\left(2 \lVert \chib^{R_n}_{xyz} - \chib^{T_m}_{xyz} \rVert \; , \; \frac{c^2 \sigma_a^2}{8\pi^2 f_c^2 |\zeta^{nm}|^2} \right) \label{eqn:range_measurement}
\end{align}
We denote all the radar parameters, signal parameters, and noise parameters as a constant $\Gamma=\frac{\sigma_a^2 \pi L}{2 P_t \Lambda_t \Lambda_r \Xi} $. For brevity, we define shorthand notation  as 
$
\mu_{z^{nm}} = 2 \lVert \chib^{R_n}_{xyz} - \chib^{T_m}_{xyz} \rVert $ and $\sigma_{z^{nm}} = \Gamma \norm{\chib^{R_n}_{xyz} - \chib^{T_m}_{xyz}}^4 $. %\frac{c^2 \sigma_a^2}{8\pi^2 f_c^2 |\zeta^{nm}|^2}$.

Existing research has explored optimal radar placement when the covariance matrix of the range measurement depends on the radar-target geometry \cite{moreno2011optimal,yan2017exact,he2015generalized}, primarily focusing on stationary targets and radars. Given that the range measurement model relies on the positions of both the radars and targets, the next section describes the target and radar kinematic models. This will facilitate the definition of the targets' conditional transition probability in the targets' state space model.

\vspace{-1em}
\section{Process Model}
\label{sec:process_modeling}
We describe the kinematic model of a target to finalize the probabilistic state space model for the concatenated state of all the targets, encompassing both a conditional transition probability and the conditional measurement probability described in \ref{sec:radar_signal_model}.

\vspace{-1em}
\subsection{Target Kinematics: Constant Velocity Model}
The target kinematic model for a single target follows constant velocity motion with acceleration noise \cite{baisa2020derivation,blackman1999design}, which specifies the single-target transition model:
\begin{align}
\chib^T_{k+1} &= A_{\text{single}}\chib_k^T + \epsilon_w \label{eqn:dynamics_single_target} 
\end{align}
where 
\begin{align}
    A_{\text{single}} &= \begin{bmatrix} \Ib_{3\times3} & \Delta t  ~ \Ib_{3\times3} \\ \mathbf{0}_{3\times3} & \Ib_{3\times3}
    \end{bmatrix} 
\end{align}
and $\epsilon_w$ is the acceleration noise which follows a zero  mean Gaussian \gls{pdf} with covariance
\begin{align}
    %      \Wb_{\text{single}} &= \begin{bmatrix}
    %     \frac{\Delta t^4}{4} & 0 & 0 & \frac{\Delta t^3}{2} & 0 & 0 \\ 
    %     0 & \frac{\Delta t^4}{4} & 0 & 0 & \frac{\Delta t^3}{2} & 0 \\ 
    %     0 & 0 & \frac{\Delta t^4}{4} & 0 & 0 & \frac{\Delta t^3}{2}\\ 
    %     \frac{\Delta t^3}{2} & 0 & 0 & \Delta t^2 & 0 & 0 \\
    %     0 & \frac{\Delta t^3}{2} & 0 & 0 & \Delta t^2 & 0 \\
    %     0 & 0 & \frac{\Delta t^3}{2} & 0 & 0 & \Delta t^2\\
    % \end{bmatrix} \cdot \sigma_W^2 \\
    \Wb_{\text{single}} &= \Wb_{\Delta t} \mathbf{\Sigma}_w \Wb_{\Delta t}^T 
\end{align}
where 
\begin{align}
    \Wb_{\Delta t} = \begin{bmatrix}
        \frac{1}{2} \Delta t^2 ~ \Ib_{3\times3} \\
        ~ \Delta t ~~ \Ib_{3\times3}
    \end{bmatrix} ~ , ~     \Sigma_w &= \Ib_{3x3} ~ \sigma_W^2 
\end{align}
and 
$\sigma_W$ determines the degree of deviation of the dynamics from constant velocity via the strength of the acceleration variance.
Thus, the targets' transition model (for all targets concatenated as a column vector) is:
\begin{align}
     \chib^{T_{1:M}}_{k+1} &= A \chib^{T_{1:M}}_k + \epsilon_{W} \label{eqn:transition_model} \\
    A &= \bm{I}_{M} \otimes A_{\text{single}} \quad W = \bm{I}_{M} \otimes W_{\text{single}} 
\end{align}
where $I_{M}\in \Re^{M\times M}$ is the identity matrix. Similarly, the measurements of all radar-target pairs may be concatenated to measurement measurement model of the targets' state-space equations (based on \cref{eqn:range_measurement}):
\begin{align}
\mathbf{z}^{T_{1:M}} &= \begin{bmatrix}
\mu_{z^{11}} & \ldots & \mu_{z^{1M}} \\
\mu_{z^{21}} & \ldots & \mu_{z^{2M}} \\
\vdots & \ddots & \vdots \\
\mu_{z^{N1}} & \ldots & \mu_{z^{NM}} \\
\end{bmatrix} +  \begin{bmatrix}
\sigma_{z^{11}} & \ldots & \sigma_{z^{1M}} \\
\sigma_{z^{21}}  & \ldots & \sigma_{z^{2M}}  \\
\vdots & \ddots & \vdots \\
\sigma_{z^{N1}} & \ldots & \sigma_{z^{NM}}  \\
\end{bmatrix} \odot \epsilon_\mathbf{z} \label{eqn:measurement_prob} \\
&= \mathbf{H(\chib^{T_{1:M}}_{xyz}}) + \Omega \odot \epsilon_\mathbf{z}
\end{align}
where $\sigma_{z^{nm}}$ is the standard deviation of the range measurement for radar $n$ and target $m$, as defined by the variance in \cref{eqn:range_measurement}, while $\epsilon_\mathbf{z}$ is standard normal Gaussian \gls{pdf}. Therefore the final state-space equations for the targets is shown in \cref{eqn:state_space_meas,eqn:state_space_tran}:
\begin{align}
     \chib^{T_{1:M}}_{k+1} &= A\chib^{T_{1:M}}_k + \epsilon_{W} \label{eqn:state_space_tran} \\
    \mathbf{z}^{T_{1:M}}_{k+1} &= \mathbf{H}\left((\chib^{T_{1:M}}_{xyz})_{k+1}\right) + \Omega \odot \epsilon_\mathbf{z} \label{eqn:state_space_meas} 
\end{align}
Given our focus on dynamic targets and dynamic radars, it's essential to outline a basic kinematic model where the radar position is determined by control inputs and adheres to specific control and dynamic constraints. Thus, we next define the second-order unicycle kinematic model for the radars.

\vspace{-1em}
\subsection{Radar Kinematics: Second-Order Unicycle Model}
Each radar follows the second-order unicycle model. We use the Euler discretization of the unicycle continuous dynamic model with heading acceleration and angular acceleration control inputs $u = [u_{a},u_{\dot{\omega}}]^T$ to define the discrete time kinematic model:
% x_{k+1} &= x_{k} + \dot{x}_k \cos{\theta_k} \Delta t + \frac{1}{2} (u_{a})_k \cos{\theta_k} \Delta t^2  \\
% y_{k+1} &= y_{k} + \dot{y}_k \sin{\theta_k} \Delta t + \frac{1}{2} 
% (u_{a})_k \sin{\theta_k} \Delta t^2 \\
% z_{k+1} &= z_{k} \\
% \theta_{k+1} &= \theta_{k} + \omega_k \Delta t + \frac{1}{2} u_{\dot{\omega}_k} \Delta t^2 \\
% \vvec_{k+1} &= \vvec_k + (u_{a})_k \Delta t \\
% \omega_{k+1} &= \omega_{k} + (u_{\dot{\omega}})_k \Delta t \\
\begin{align}
    \chib^{R}_{k+1} &= \chib^{R}_k + G_k(\chib^{R}_k,u_k) \label{eqn:radar_kinematics}
\end{align}
with control, velocity, and angular velocity limits,
\begin{equation}
\begin{aligned}
\begin{split}
\underbar{$u$}_a \leq u_{a_k} &\leq \bar{u}_a \; \forall k \\
\underbar{$u$}_{\dot{\omega}} \leq u_{\dot{\omega}_k} &\leq \bar{u}_{\dot{\omega}} \; \forall k  \\
\end{split} \quad
\begin{split}
\underbar{$v$} \leq v_k &\leq \bar{v} \; \forall k \\
\underbar{$\omega$} \leq \omega_k &\leq \bar{ \omega } \; \forall k. 
\end{split}
\end{aligned}
\label{eqn:radar_kinematics_constrained}
\end{equation}
With the radars' positions defined relative to control inputs, the following section outlines the derivation of the \glspl{fim} with respect to the targets' positions. This is based on the radar-target measurement model outlined in  \cref{eqn:range_measurement,eqn:measurement_prob}, and the radars' positions parameterized by \cref{eqn:radar_kinematics,eqn:radar_kinematics_constrained}.

\vspace{-1em}
\section{Fisher Information Matrix}
\label{sec:fim_definition}

The \gls{sfim} measures the amount of information that an observable random variable $z$ (range measurements) carries about an unknown parameter $\theta$ (targets' states) \cite{tichavsky1998posterior,kay1993fundamentals}:

\begin{align}
    \Jb_D(\theta) = E\left[\left( \nabla_{\theta} \log f\left(z|\theta \right) \right) \left( \nabla_{\theta} \log f\left(z|\theta \right) \right)^T\right]
\end{align}
Under the assumption of independence, the range measurements likelihood in \cref{eqn:measurement_prob} is represented as:
\begin{align}
    f\left(\mathbf{z}^{T_{1:M}}|\chib^{T_{1:M}}_{xyz} ; \chib^{R_{1:N}}_{xyz} \right) &= \prod_{n=1}^{N} \prod_{m=1}^{M} f(z^{nm} | \chib^{T_m}_{xyz} ; \chib^{R_n}_{xyz})
\end{align}
Given that the measurement is a Gaussian \gls{pdf}, with the mean and covariance parameterized by the targets' positions, and that information of measurements is additive by the independence assumption, we use the derivation of general Gaussian \gls{fim} \cite{kay1993statistical} to calculate the \gls{sfim} for a multiple targets' positions with multiple radars:
\begin{align}
    \Jb(\chib^{T_m}_{xyz};\chib^{R_n}_{xyz}) &= (\chib^{T_m}_{xyz} - \chib^{R_n}_{xyz})(\chib^{T_m}_{xyz} - \chib^{R_n}_{xyz})^T \times \nonumber \\
    &\left( \frac{4}{\Gamma \lVert\chib^{R}_{xyz} - \chib^{T}_{xyz} \rVert^6} + \frac{8}{\lVert \chib^{R}_{xyz} - \chib^{T}_{xyz} \rVert^4} \right) \label{eqn:fim_single_radar_single_target} \\
    \Jb(\chib^{T_m}_{xyz};\chib^{R_{1:N}}_{xyz}) &= \sum_{n=1}^N \Jb(\chib^{T_m}_{xyz} ; \chib^{R_n}_{xyz})     \label{eqn:fim_multi_radar_single_target} 
\end{align}
\vspace{-2em}
\begin{align}
    &\Jb(\chib^{T_{1:m}}_{xyz};\chib^{R_{1:n}}_{xyz}) = \nonumber \\
        &\begin{bmatrix}
        \Jb(\chib^{T_{1}}_{xyz};\chib^{R_{1:N}}_{xyz}) & \!\! 0 & \!\!\ldots & \!\!0 \\
        0 & \!\! \Jb(\chib^{T_{2}}_{xyz};\chib^{R_{1:N}}_{xyz}) & \!\! \ldots & \!\! 0 \\
        \vdots & \!\! \vdots & \!\! \ddots & \!\! \vdots \\
        0 & \!\! 0 & \!\! \ldots & \!\! \Jb(\chib^{T_{M}}_{xyz};\chib^{R_{1:N}}_{xyz})
    \end{bmatrix} \label{eqn:fim_multi_radar_multi_target}
\end{align}
Equation \cref{eqn:fim_single_radar_single_target} is the \gls{sfim} of a single target position parameterized by a single radar and \cref{eqn:fim_multi_radar_single_target} is the \gls{sfim} of a single target positions parameterized by multiple radars.

When there is prior knowledge about the dynamics of the target state (\cref{eqn:dynamics_single_target}), then this information may be incorporated into the \gls{sfim} to form the \gls{pfim}:
\begin{align}
    \Jb = \Jb_D + \Jb_P \\
    \Jb_P = E[\Delta_{\chib^T}^{\chib^T} \log f(\chib^T)] \label{eqn:pfim_single}
\end{align}
Here, $\Jb_D$ represents the information about the unknown target state obtained from the data, while $\Jb_P$ reflects the information about the unknown target state based on prior assumptions. Now, considering that the targets' state includes velocity, we update \cref{eqn:fim_multi_radar_multi_target} to be $\Jb_D$ for the \gls{pfim}, with the block matrices as:
\begin{align}
&\Jb(\chib^{T_n};\chib^{R_{1:N}}_{xyz}) =  \begin{bmatrix}
    \Jb(\chib^{T_n}_{xyz};\chib^{R_{1:N}}_{xyz}) & 0 \\
    0 & 0
\end{bmatrix}  \label{eqn:fim_multi_radar_multi_target_velocity} 
\end{align}

For discrete time filtering systems, such as \cref{eqn:state_space_meas,eqn:state_space_tran}, the
\cref{eqn:pfim_single} is recursively computed at time $k+1$ from time step $k$ using the famous recursion updates in \cite{tichavsky1998posterior}. The recursion equations are written as 
\begin{align}
    &\Jb_{k+1} (\chib^{T_{1:M}}_{k+1}) = \Db_k^{22} - \Db_k^{21} (\Jb_k + \Db_k^{11})^{-1} \Db_k^{12} \label{eqn:pfim_recursion}
\end{align}
where
\begin{align}
    &\Db_k^{11} = E[-\Delta_{\chib^{T_{1:M}}_k}^{\chib^{T_{1:M}}_k} \log p(\chib^{T_{1:M}}_{k+1} | \chib^{T_{1:M}}_k)]  \\
    &\Db_k^{12} = E[-\Delta_{\chib^{T_{1:M}}_k}^{\chib^{T_{1:M}}_{k+1}} \log p(\chib_{k+1} | \chib_k)] = [\Db_k^{21}]^T \\
    &\Db_k^{22} = E\left[-\Delta_{\chib^{T_{1:M}}_{k+1}}^{\chib^{T_{1:M}}_{k+1}} \log p(\chib^{T_{1:M}}_{k+1} | \chib^{T_{1:M}}_{k}) \right] + \nonumber \\
    & \qquad E\left[-\Delta_{\chib^{T_{1:M}}_{k+1}}^{\chib^{T_{1:M}}_{k+1}} \log p(z_{k+1} | \chib^{T_{1:M}}_{k+1}) \right] 
\end{align}
Each $D$ term in equation \cref{eqn:pfim_recursion} is (\gls{pdf} regularity assumptions) as 
\begin{align}
    &\Db_k^{11} =\Ab^T \Wb^{-1} \Ab  \label{eqn:d11}\\
    &\Db_k^{12} = - \Ab^T \Wb^{-1} \label{eqn:d12} \\
    &\Db_k^{22} =  \Wb^{-1} + E_{f(\chib^{T_{1:M}}_{1:k+1})} \left[\Jb_{\Db_k}(\chib^{T_{1:M}}_{k+1})) \right] \label{eqn:d22}
\end{align}
where due to the Markov assumption on the transition model in the targets' state space equation (\cref{eqn:transition_model}), the joint \gls{pdf} $f(\chib^{T_{1:M}}_{1:k})$ is simplified as
\begin{align}
    f(\chib^{T_{1:M}}_{1:k}) = f(\chib^{T_{1:M}}_{0}) \prod_{t=1}^{k+1} f(\chib^{T_{1:M}}_{t} | \chib^{T_{1:M}}_{t-1}) \label{eqn:markov_simplified}
\end{align}
After substituting equations \cref{eqn:d11,eqn:d12,eqn:d22} into \cref{eqn:pfim_recursion}, and leveraging the Woodbury matrix identity \cite{bishop2006pattern}, we arrive at  \cite{simplifed_pfim_recursion}:
\begin{align}
    \Jb_{k+1} (\chib^{T_{1:M}}_{k+1}) &= (\Wb + \Ab^T \Jb_k^{-1} \Ab)^{-1} + \notag \\ & \quad  \quad E_{f(\chib^{T_{1:M}}_{1:k+1})} \left[\Jb_{\Db_k}(\chib^{T_{1:M}}_{k+1})) \right] \label{eqn:pfim_simplified}
\end{align}
\cref{eqn:pfim_simplified} is preferred over \cref{eqn:pfim_recursion} for numerical stability. This preference arises because it avoids the inversion of the $W$ matrix, which can be close to singular in cases where the conditional transition PDF is described by constant velocity models.

To compute the expectation over the target state for \cref{eqn:d22}, as well as objective terms involving \cref{eqn:fim_multi_radar_multi_target} or any other objective term in the \gls{mpc} objective that involves the targets' states, we employ the \gls{ckf} to describe \cref{eqn:markov_simplified}.

\glspl{ckf} efficiently leverage third-order spherical cubature integration into both the prediction and correction steps of an additive Gaussian noise Kalman filter \cite{sarkka2023bayesian}. To compute the expectations described in the previous paragraph, we propagate the sigma points of the \gls{ckf} multiple time steps into the future, employing a number of sigma points equal to twice the dimension of the state variable. Compared to a \gls{pf} \cite{sarkka2023bayesian}, the \gls{ckf} has lower computational complexity, as the number of evaluation points scales linearly with the state size, while the number of particles in a \gls{pf} may substantially increase.

With our current formulations in \cref{sec:radar_signal_model,sec:process_modeling,sec:fim_definition}, we formulate a discrete-time finite-horizon \gls{mpc} objective  using dynamic, control, and mission-aware constraints.

\vspace{-1em}
\section{Model Predictive Controller Objective}
\label{sec:mpc_objective}
\gls{mpc} is an optimal control technique that determines the control actuation minimizing a cost function for a constrained dynamical system over a finite, receding horizon \cite{morari1999model,khaled2018practical}. At each time step of the receding horizon objective, we enforce minimum distances between radars and targets to prevent potential destruction by adversaries using soft penalties. Similarly, we ensure safe separation distances between mobile radars to prevent collisions. Additionally, we incorporate the radar's acceleration, angular acceleration, angular speed, and heading velocity constraints as described in \cref{eqn:radar_kinematics_constrained} into the dynamic equation using clipping functions. Finally, we directly integrate the radar's dynamic equation (\cref{eqn:radar_kinematics_constrained}) into the objective using direct shooting \cite{directshooting}.

As a result, converting hard constraints to soft/implicit constraints results in an unconstrained, discontinuous, and non-differentiable optimization formulation. Lastly, we incorporate the uncertainty of the targets' state by taking expectation of the \gls{mpc} objective function with respect to the \gls{ckf} estimator of the targets' state:
{\hspace{-5cm}
\begin{align}
    &S(\uvec_{k:k+K+1} ; \chib^{T_{1:M}}_{k:k+K+1}) \!=\!  E_{\tilde{f}} \bigr[S_{traj}(\uvec_{k:k+K+1} ; \chib^{T_{1:M}}_{k:k+K+1})  \notag \\
    & \quad \quad \quad  \quad \quad \quad \quad  + \alpha_1 S_{R2R}(\uvec_{k:k+K}) \notag \\
    & \quad \quad \quad  \quad \quad \quad \quad + \alpha_2 S_{R2T}(\uvec_{k:k+K} ; \chib^{T_{1:M}}_{k:k+K}) \bigr] \label{eqn:mpc_obj_unconstrained}  \\
    &\text{where } \notag\\
    & S_{traj}\left(\chib^{R_{1:N}}_{k:k+K+1} ; \chib^{T_{1:M}}_{k:k+K+1}\right) = \notag \\
    &\quad \sum_{t=k}^{k+K+1} \gamma^{t-k} \log \det \left(J(\chib_{t}^{T_{1:M}} ; \chib_{t}^{R_{1:M}} \right)  \notag   \\
    & S_{R2T}(\uvec_{k:k+K+1} \!) \! = \notag \\
    &\quad \; \; \!\! \sum_{t=k}^{k+K+1} \!\! \gamma^{t-k} \!\! \sum_{m=1}^M \! \sum_{n=1}^N \! \mathbf{1} \! \left( \lVert \chib_t^{T_n} \! - \! \chib_t^{R_m}(\uvec_{t})  \rVert \! \leq \! R_{R2T} \! \right) \notag \\
    & S_{R2R}(\uvec_{k:k+K+1}) = \notag \\
    & \quad \; \sum_{t=k}^{k+K+1} \gamma^{t-k} \!\!\!\!\!\! \sum_{\{i<j: i,j \in M\}} \!\!\!\!\!\! \mathbf{1} \left( \lVert \chib_t^{R_i}(\uvec_{t}) - \chib_t^{R_j}(\uvec_{t})  \rVert \leq R_{R2R} \right) 
\end{align} 
where $\alpha_1,\alpha_2$ are hyper parameters that determine how much to penalize the radars for violating the soft constraints $S_{R2R},S_{R2T}$ respectively, $\gamma$ is the discount factor, and $K$ is the horizon length. The \gls{ckf} estimator \gls{pdf} on the targets' state over the horizon is
\begin{align}
\tilde{f}(\chib_{k:k+K+1}^{T_{1:M}}) = f_{ckf}(\chib^{T_{1:M}}_{k}|\mathbf{z}_{1:k-1}) \! \prod_{t=k+1}^{k+K+1} \! f(\chib^{T_{1:M}}_{t} | \chib^{T_{1:M}}_{t-1})    
\end{align}
}

This \gls{mpc} problem requires an efficient solver that can handle highly nonlinear, discontinuous, and non-differentiable functions. Therefore, we employ a sampling-based strategy, \gls{mppi} \cite{williams2018information}, as discussed in the following section.

\vspace{-1em}
\subsection{Model Predictive Path Integral}
\gls{mppi} is a sampling-based \gls{mpc} algorithm designed for optimizing nonlinear stochastic systems subject to complex cost functions in real time \cite{7487277}. The assumption of \gls{mppi} is the control input is random and distributed as Gaussian random vector $\vvec \sim N(\uvec,\bm{\Sigma})$. Due to the stochasticity of the control input, rather than solving for a deterministic optimal control input we solve for an optimal control \gls{pdf} $Q^*$ \cite{theodorou2010generalized}: 
\begin{align}
    Q^* = \frac{1}{\eta} \exp\left(-\frac{1}{b}S(\vvec_{k:k+K+1}) \right) P(\vvec_{k:k+K+1})
\end{align}
where $P(\cdot)$ is the \gls{pdf} of an control input sequence in an uncontrolled system ($\uvec_{k:k+K+1} = 0)$, $b$ is the temperature of the distribution, and $\eta$ is the evidence.
However, the optimal control \gls{pdf} likely does not possess an analytical form. Therefore, by minimizing the \gls{kbl} between $Q^*$ and a variational Gaussian \gls{pdf} $Q$ parameterized by $u,\Sigma$ \cite{williams2018information} with respect to $u$ gives the optimal control at $k$ as \cite{williams2018information} 
\begin{align}
    \uvec^*_k = \int q^*(\bigvvec) \vvec_k d\bigvvec \label{eqn:optimal_control_mppi}
\end{align}

Since we cannot express $Q^*$ analytically in \cref{eqn:optimal_control_mppi}, we leverage \gls{ais} to approximate the expectation as follows \cite{asmar2023model,williams2018information}: 
\begin{enumerate}
    \item sample control trajectories $\vvec_{k:k+K+1}^{(i)}$ from distribution $Q$ with mean $\uvec_{k:k+K+1}$ and covariance $\Sigma$
    \item evaluate the objective function for each control trajectory sample $S(\vvec_{k:k+K+1}^{(i)})$
    \item compute the importance sampling corrective weights for each sample $\tilde{w}^{(i)} \propto S(\vvec^{(i)})$
    \item adapt the mean and covariance of the \gls{pdf} $Q$ 
    \item repeat steps 1-4
\end{enumerate}

Due to the high dimensionality of the targets' control input over a horizon and the limited number of samples in \gls{mppi}, the sample covariance matrix for control inputs can overfit, leading to inaccurate variances \cite{theiler2012incredible}. This often results in poor control inputs from the proposal distribution of \gls{mppi}. To address this, we adopt the \gls{lw} for \gls{ais}, which approximates \gls{oas} shrinkage and helps mitigate overfitting, following the method described in \cite{asmar2023model} \footnote{The authors in \cite{asmar2023model} do not explicitly discuss Oracle Approximating Shrinkage (OAS) or Ledoit-Wolfe, but their publicly available code demonstrates methods for estimating the sample covariance, including Ledoit \& Wolf, Schaffer \& Strimmer, Rao-Blackwell, and OAS. A reference in the publicly available code is \href{https://github.com/sisl/MPOPIS/wiki/MPOPIS-Details}{https://github.com/sisl/MPOPIS/wiki/MPOPIS-Details}.}.

\begin{figure*}[h!]

\centering
\subfigure[Proposed approach: Each radar moves towards the targets.]{\includegraphics[width=8cm]{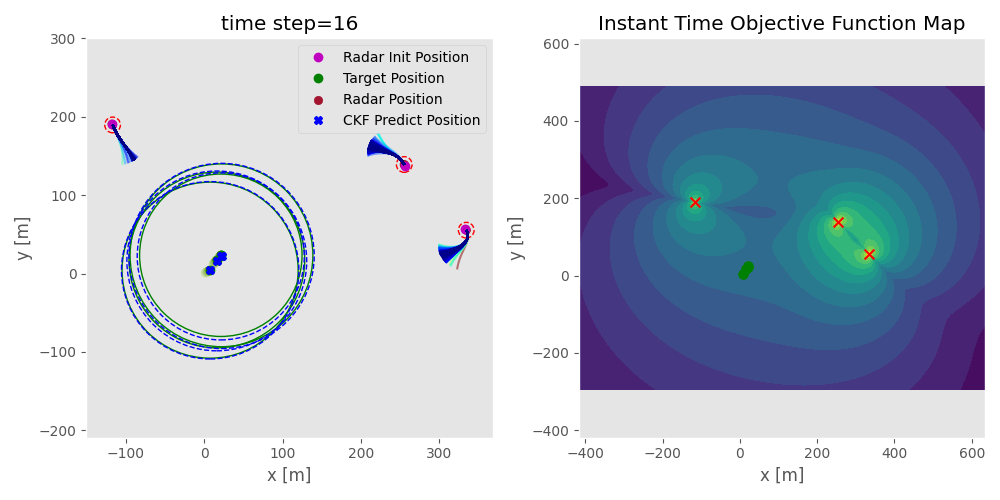}}\hfil
\subfigure[\cite{hung2020range,bishop2010optimality,crasta2018multiple} range model: The upper right radar moves away from targets]{\includegraphics[width=8cm]{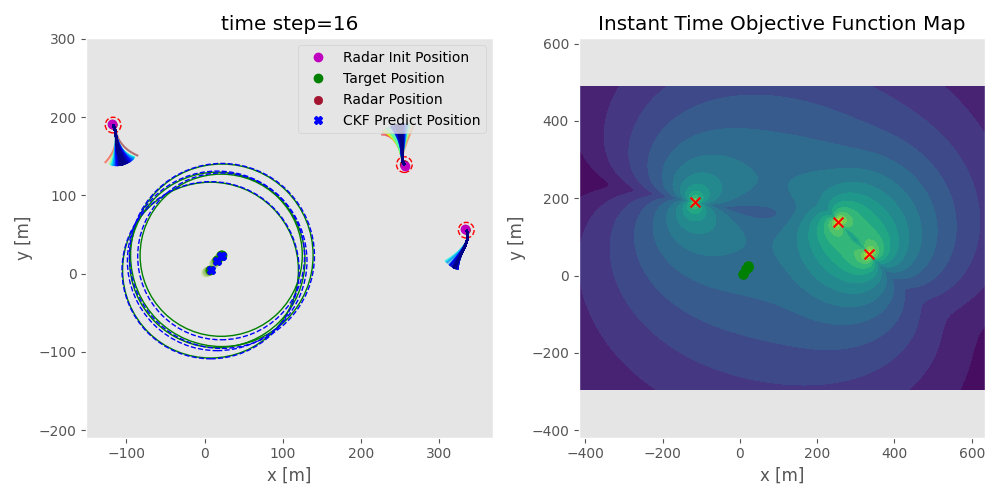}}\hfil 
\vspace{-.5em}
\subfigure[Radars move toward targets as closely as feasible.]{\includegraphics[width=8cm]{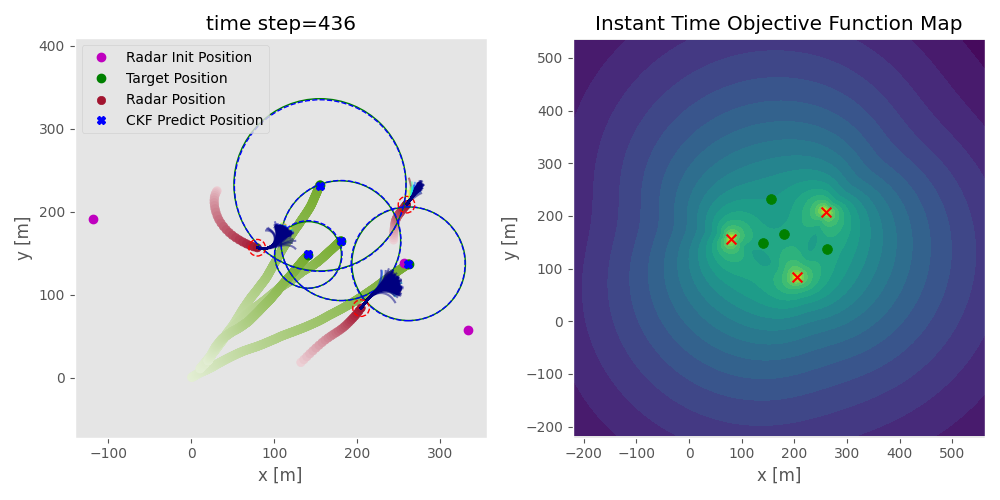}} \hfil 
\subfigure[Radars encircle targets, but the bottom most radar remains distant.]{\includegraphics[width=8cm]{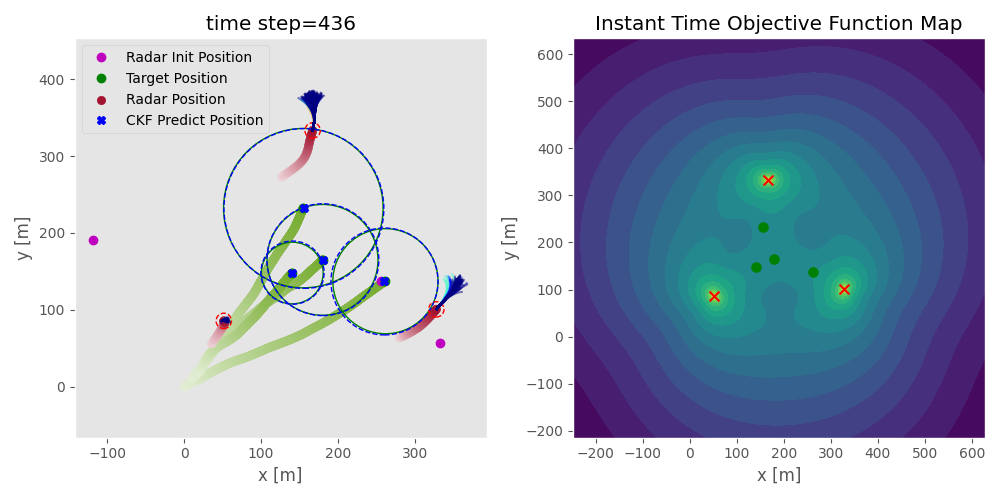}}\hfil  
\vspace{-.5em}
\subfigure[Radars maintain circular formation with targets circle formation.]{\includegraphics[width=8cm]{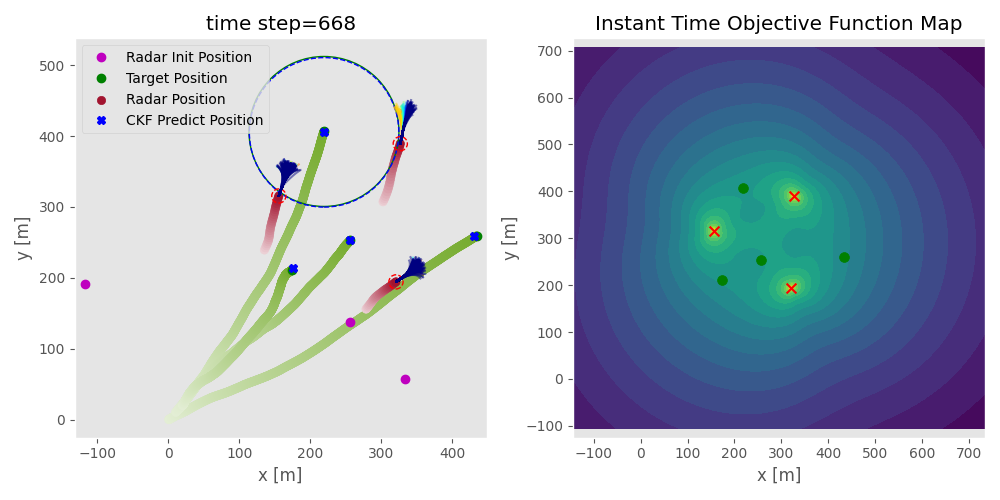}} \hfil 
\subfigure[Radars encircle the targets, but bottom most radar moves away from targets.]{\includegraphics[width=8cm]{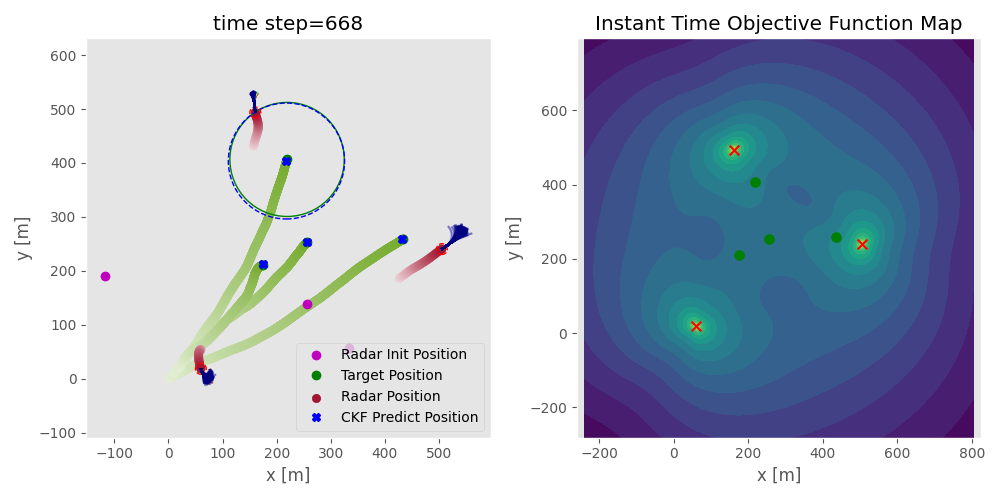}}\hfil  
\vspace{-.5em}
\subfigure[Radars maintain a circular formation within the circle of targets.]{\includegraphics[width=8cm]{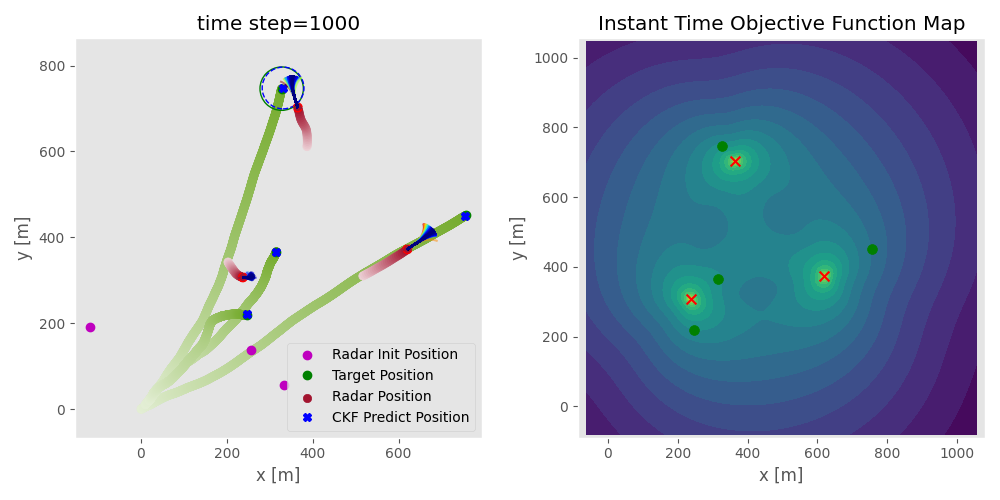}} \hfil 
\subfigure[Radars maintain encirclement formation around targets cricle]{\includegraphics[width=8cm]{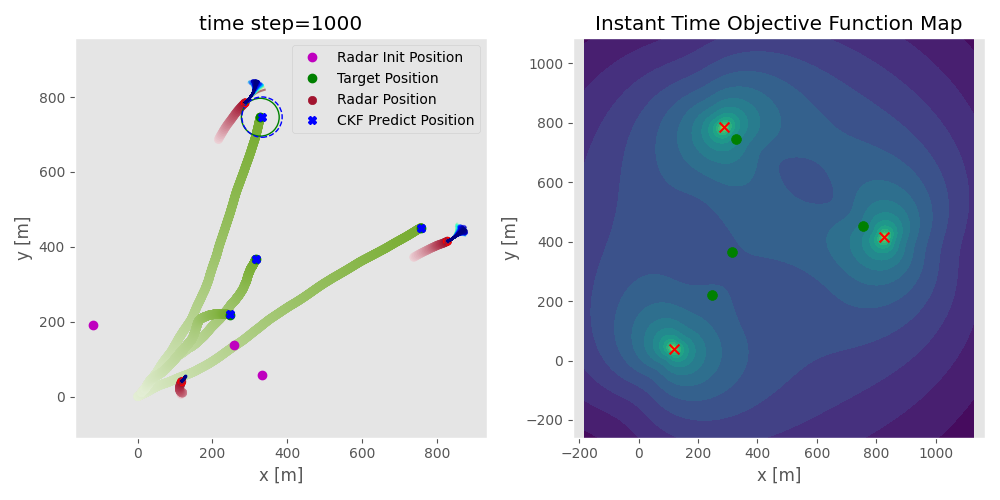}}\hfil  
\vspace{-.5em}
\caption{ \small 
The left column shows snapshots of target localization using a range measurement model with covariance dependent on radar-target distance \cite{godrich2010target}, while the right column depicts tracking with a range measurement model employing constant diagonal covariance \cite{bishop2010optimality}. In the first three rows, each subfigure's left panel shows the 3D projection onto 2D of continuously optimizing radar placement at a specific time. The colored lines emanating from the radars depict the weighted MPPI-planned trajectories, with blue indicating better paths and red indicating worse ones. Dashed red and green circles mark the radar's (10 [m]) and the target's (125 [m]) collision avoidance boundaries, respectively, while blue dashed circles represent the targets' collision avoidance boundary based on the CKF position estimate. The radar's trajectory is the last 25 time steps. The right panel displays the log determinant of the FIM (for the ddr measurement model) concerning the 2D target position for the specific time step, meshed over the free space area, and evaluated at each mesh point. }
\label{fig:noise_compare}
\end{figure*}

\begin{figure*}[htp!]
\centering
\subfigure[3 Radar - 4 Target: RMSE with 90\% \gls{hdi}]{\includegraphics[width=5.5cm]{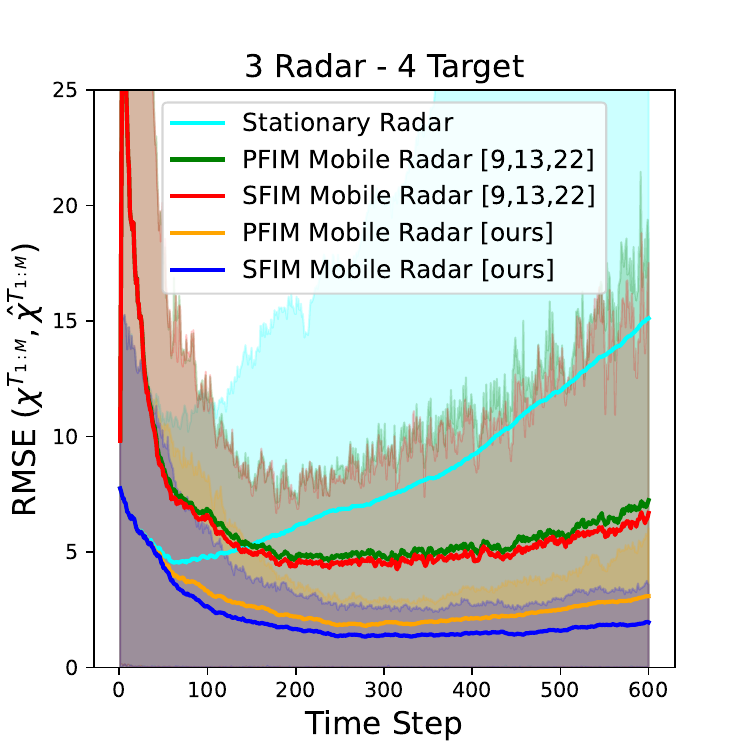}} \hfil \hfil
\subfigure[3 Radar - 4 Target: RMSE without stationary radar or \gls{hdi}]{\includegraphics[width=5.5cm]{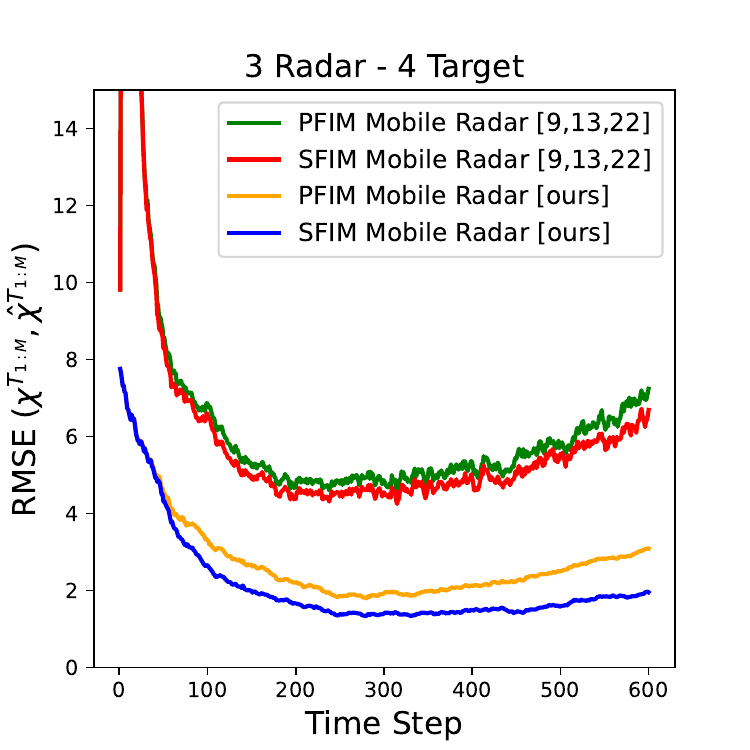}} \hfil \hfil
\subfigure[3 Radar - 4 Target: RMSE ECDF]{\includegraphics[width=5.3cm]{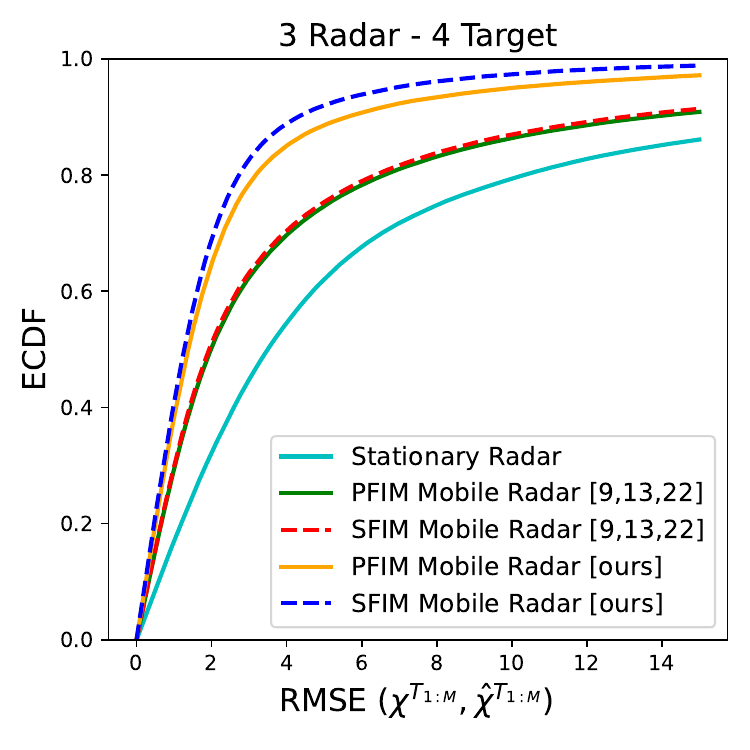}}\hfil  
\subfigure[6 Radar - 3 Target: RMSE with 90\% \gls{hdi}]{\includegraphics[width=5.5cm]{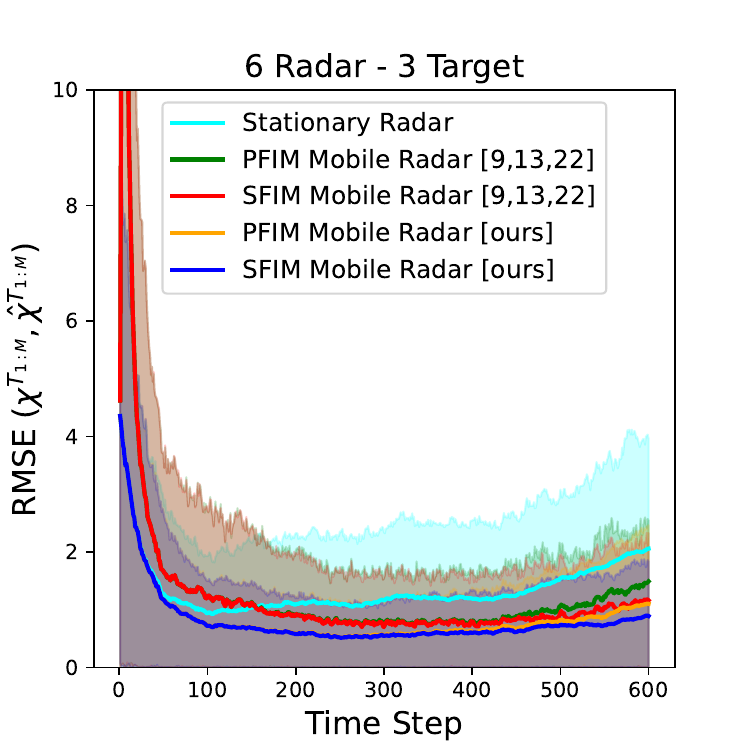}} \hfil \hfil 
\subfigure[6 Radar - 3 Target: RMSE without stationary radar or \gls{hdi}]{\includegraphics[width=5.5cm]{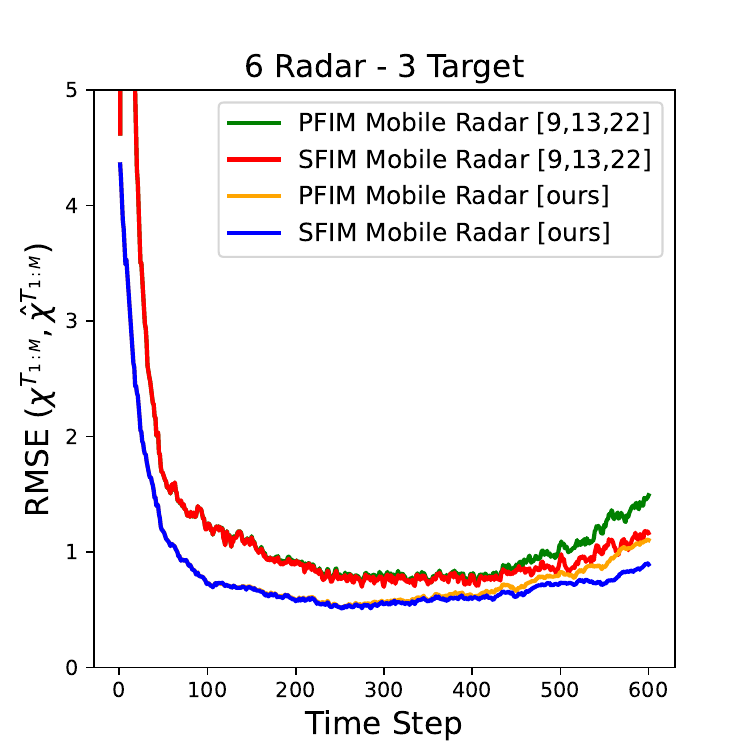}} \hfil \hfil
\subfigure[6 Radar - 3 Target: RMSE ECDF]{\includegraphics[width=5.3cm]{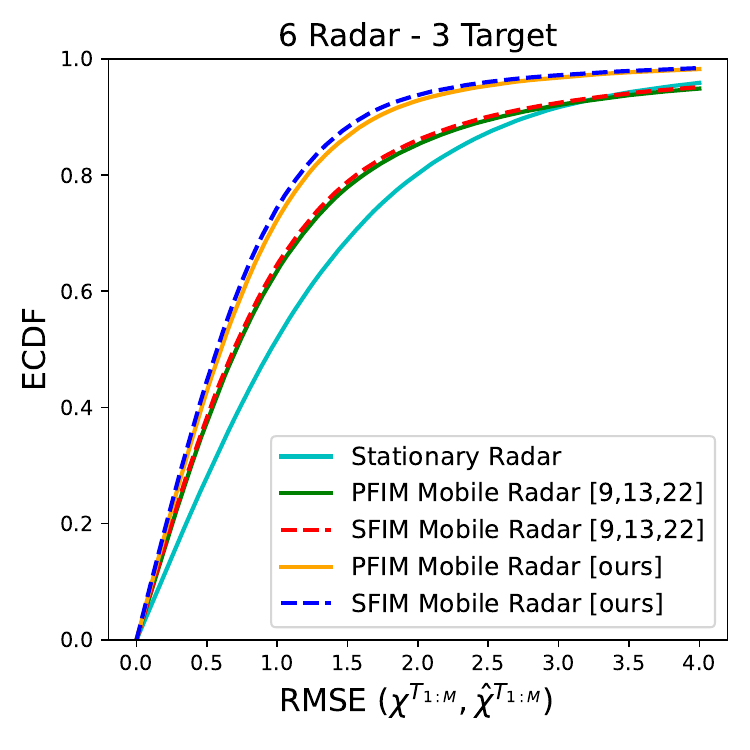}}\hfil  
\subfigure[4 Radar - 4 Target: RMSE with 90\% \gls{hdi}]{\includegraphics[width=5.5cm]{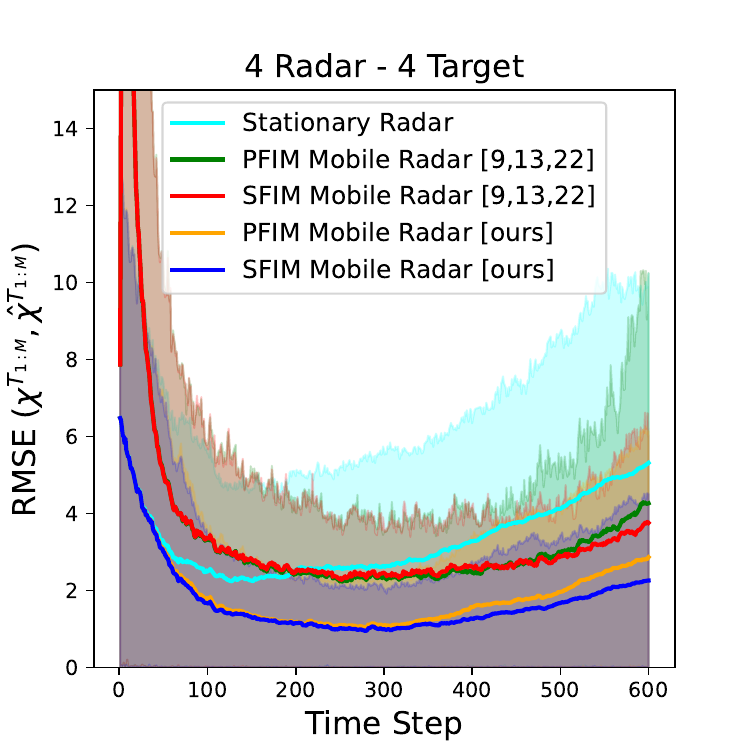}} \hfil \hfil
\subfigure[4 Radar - 4 Target: RMSE without stationary radar or \gls{hdi}]{\includegraphics[width=5.5cm]{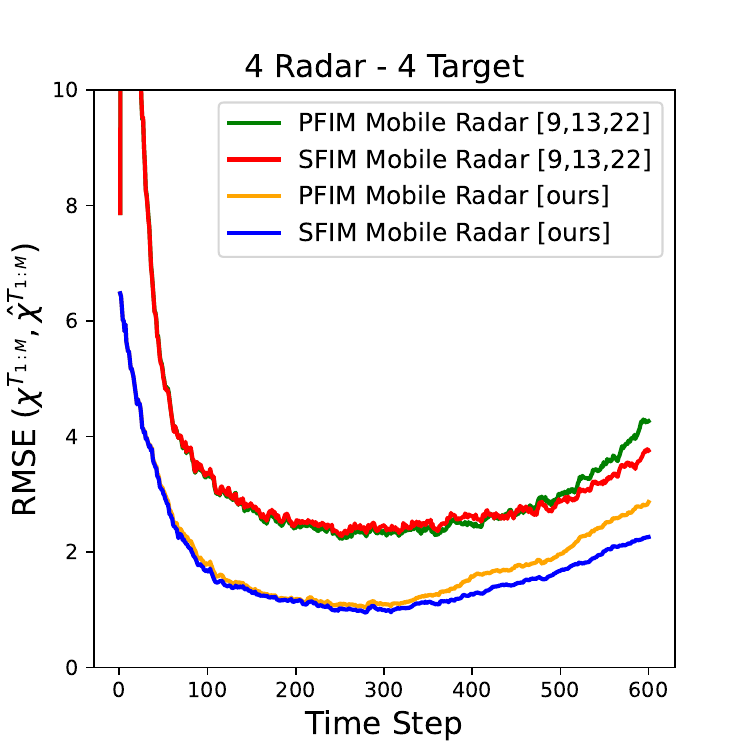}} \hfil \hfil
\subfigure[4 Radar - 4 Target: RMSE ECDF]{\includegraphics[width=5.3cm]{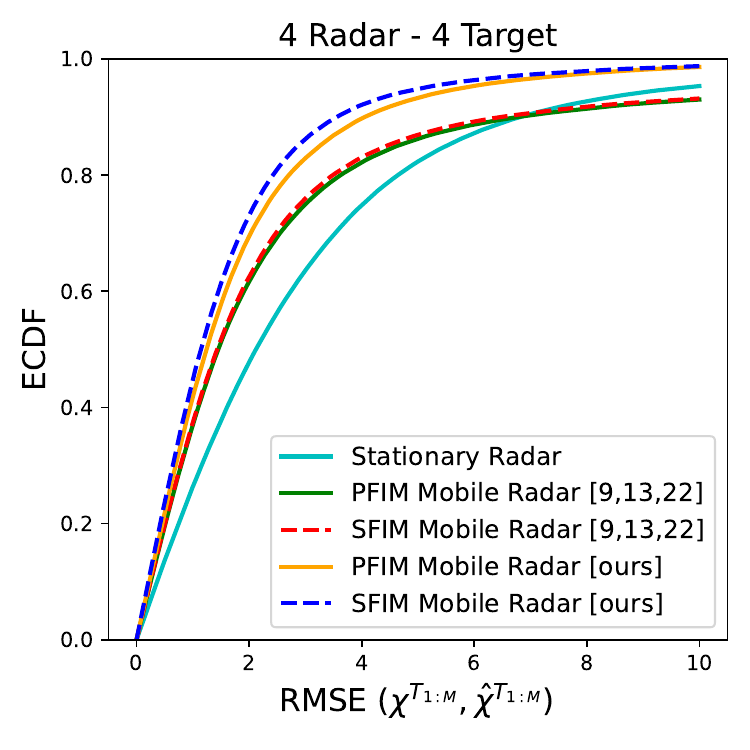}}\hfil  
\caption{ \small 
The difference in target localization errors between constant covariance and radar-based covariance for 3 radars aand 4 targets, 6 radars aand 3 targets, and 4 radars and 4 targets for the \gls{sfim} and \gls{pfim} based objective via \gls{rmse} plot with 90\% \gls{hdi} and \gls{rmse} \gls{ecdf} plots. The left column shows the mean \gls{rmse} over the targets' true state versus the CKF predicted state over 500 MC trials. The lightly shaded region is the 90\% highest density interval (HDI) of the \gls{rmse} over 500 MC trials. The middle column shows a zoom in of the left column subfigurese without stationary radar and 90\% \gls{hdi} \glspl{rmse}. The right column is the empirical \gls{ecdf} of the \gls{ckf} predicted state \gls{rmse} over time steps and MC trials. Higher \gls{ecdf} is better at a given \gls{rmse}.}
\label{fig:rmse_compare}
\end{figure*}

\RestyleAlgo{ruled}
\SetKwComment{Comment}{/* }{ */}

\begin{algorithm}[hbt!]
\caption{Continuous Optimal Radar Placement}\label{alg:contoptimalradar}
\KwData{$\chib^{R_{1:N}}_0$}
\KwResult{$\hat{\chib}^{T_{1:M}}_{1:T},\chib^{R_{1:N}}_{1:T}$}
% $y \gets 1$\;
% $X \gets x$\;
% $N \gets n$\;
\For{$k=1:T$}{
 $\chib^{T_{1:M}}_{k+1}$ = TransitionFn$\left(\chib^{T_{1:M}}_{k}\right)$ \\
 $\hat{\chib}^{T_{1:M}}_{k+1}$ = CKF.predict$\left(\hat{\chib}^{T_{1:M}}_{k}\right)$ \\
  \If{$k\% T_{control}=0$}{
    $\hat{\chib}^{T_{1:M}}_{k+1:k+K+1} \gets$ CKF.propogate$\left(\hat{\chib}^{T_{1:M}}_{k+1}\right)$\;
    $\mathbf{u}_{k:k+K+1} \gets$ MPPI$\left(\hat{\chib}^{T_{1:M}}_{k+1:k+K+1},\chib^{R_{1:M}}_k,\right)$\;
    % cost = MPC$\left(\mathbf{u}_{k:k+K+1},\chib^{R_{1:N}}_k,\hat{\chib}^{T_{1:M}}_{k+1:k+K+1}\right)$
    % \Comment*[r]{This is a comment}
    % $\mathbf{u}_{k:k+T_{control}}$ = repeat$\left(\uvec_k, T_{control}\right)$ 
    % \Comment*[r]{\tiny due to the control update frequency being smaller than the CKF frequency}
  }
  $\chib^{R_{1:N}}_{k+1} \gets $ Actuate$\left(\chib^{R_{1:N}}_{k},\mathbf{u}_k\right)$ \;
  $\mathbf{z}^{R_{1:N}}_{k+1}$ = MeasureFn$\left(\chib^{T_{1:M}}_{k+1},\chib^{R_{1:N}}_{k+1}\right)$ \;
  $\hat{\chib}^{T_{1:M}}$ = CKF.update$\left(\chib_{k+1}^{R_{1:N}},\mathbf{z}^{R_{1:N}}_{k+1}\right)$ \;
}
\end{algorithm}

The integration of the algorithms outlined in \cref{sec:radar_signal_model,sec:process_modeling,sec:fim_definition,sec:mpc_objective} forms the enhanced continuous optimal radar placement pipeline, described in Algorithm 1. This pipeline includes two key enhancements: an improved range measurement model, which significantly impacts the \gls{fim}, and a flexible and computationally efficient \gls{mpc} objective optimizer using \gls{mppi}.

%% file: appendix_paper_r1.tex
\onecolumn

\appendix
\renewcommand{\figurename}{Appendix Figure}
\renewcommand{\tablename}{Appendix Table}
% \renewcommand{\thesection}{Supplementary Section}
% \renewcommand{\thesubsection}{Supplementary Subsection}
% \renewcommand{\thesubsubsection}{Supplementary Subsubsection}
% \crefalias{section}{supp}

\setcounter{figure}{0} 
\setcounter{table}{0}
\crefalias{figure}{appfig}
\crefalias{table}{apptab}
\crefalias{section}{appsec}

% \subsection{Concept Of Operations}
% \label{appsec:conops}

% \begin{figure*}[h!]
%     \centering
%     \subfigure[Search and rescue in forest fire]{\includegraphics[width=4.5cm]{images/conops/searchandrescue.png}} \hfil 
%     \subfigure[Enemy drone localization and tracking]{\includegraphics[width=4.5cm]{images/conops/dronetracking.png}}\hfil  
%     \caption{Subfigure (a) shows the civilian application of a search and rescue mission where mobile radars navigate through a dense forest (on fire) to quickly localize and track lost victims. Subfigure (b) shows the military application of enemy drone localization and tracking in a city environment during  a cloudy day.}
%     \label{appfig:conops}
%     \vspace{-1.5em}
% \end{figure*}

\subsection{Single Simulation Realization}
\begin{figure*}[htp!]
\centering
\subfigure[Proposed approach: Each radar moves towards target]{\includegraphics[width=8.5cm]{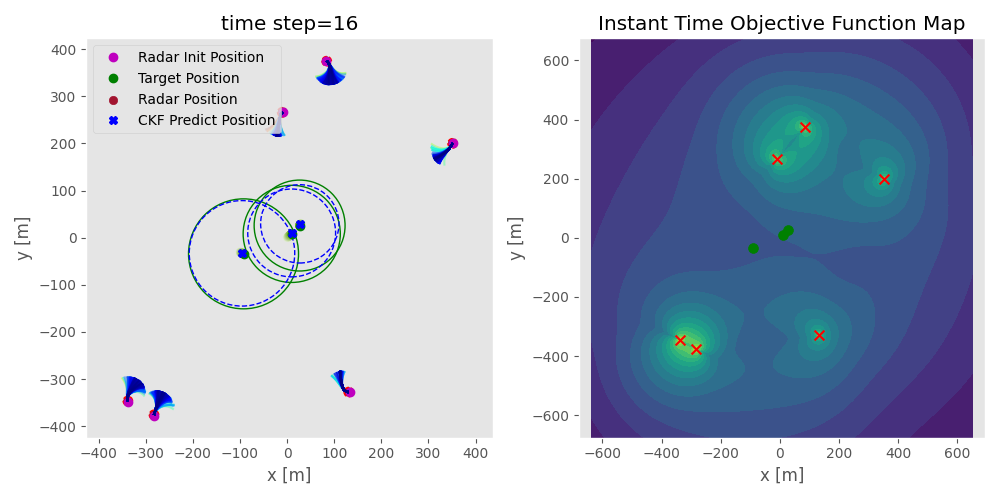}}\hfil
\subfigure[\cite{crasta2018multiple,bishop2010optimality,hung2020range} range model: Radars move to encircle targets]{\includegraphics[width=8.5cm]{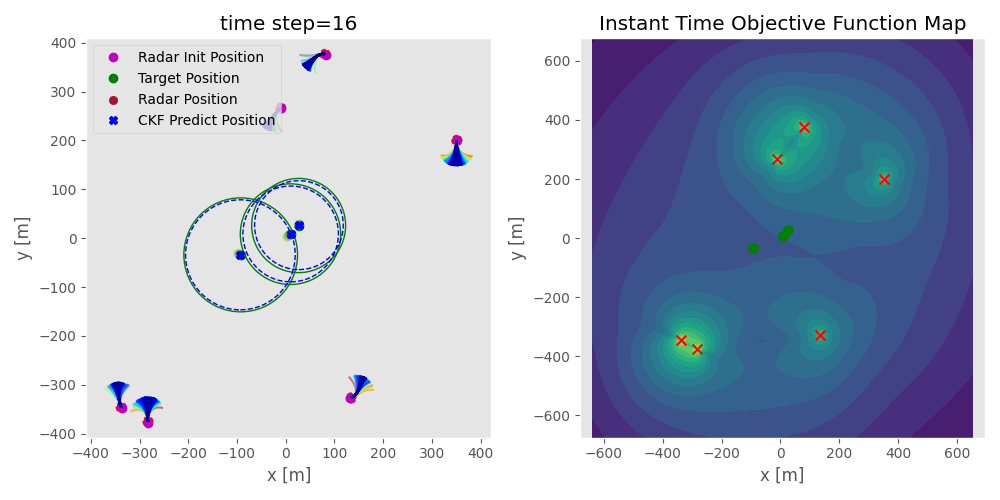}}\hfil 
\subfigure[Radars move toward targets as closely as feasible]{\includegraphics[width=8.5cm]{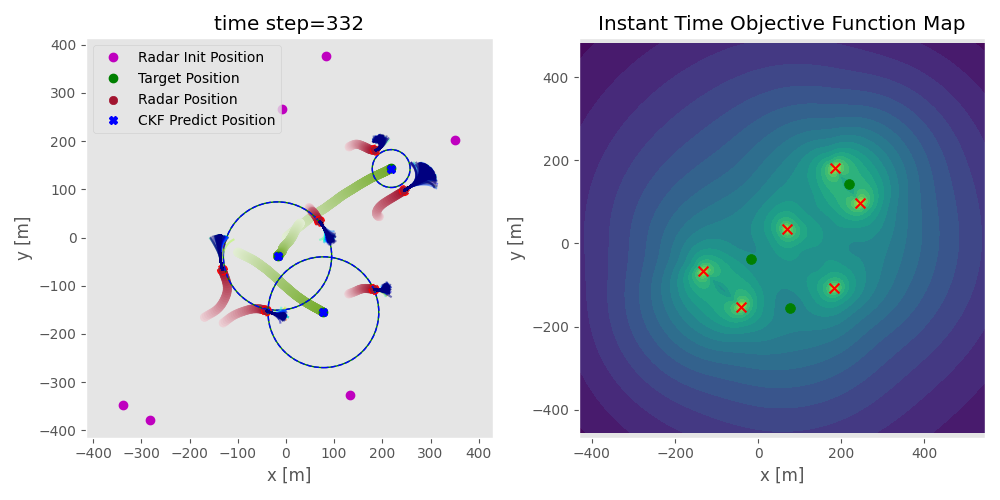}} \hfil 
\vspace{-.5em}
\subfigure[Radars encircle targets]{\includegraphics[width=8.5cm]{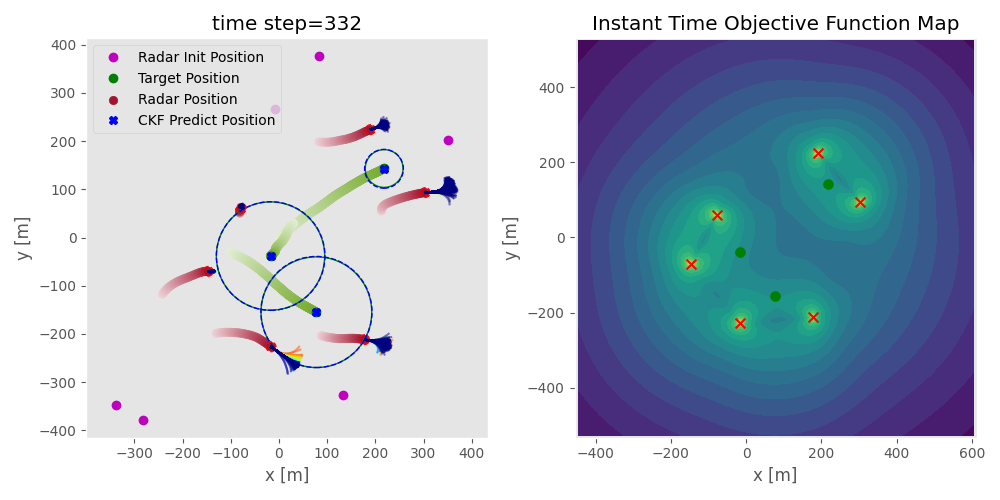}}\hfil  
\vspace{-.5em}
\subfigure[Two radars follow each target individually]{\includegraphics[width=8cm]{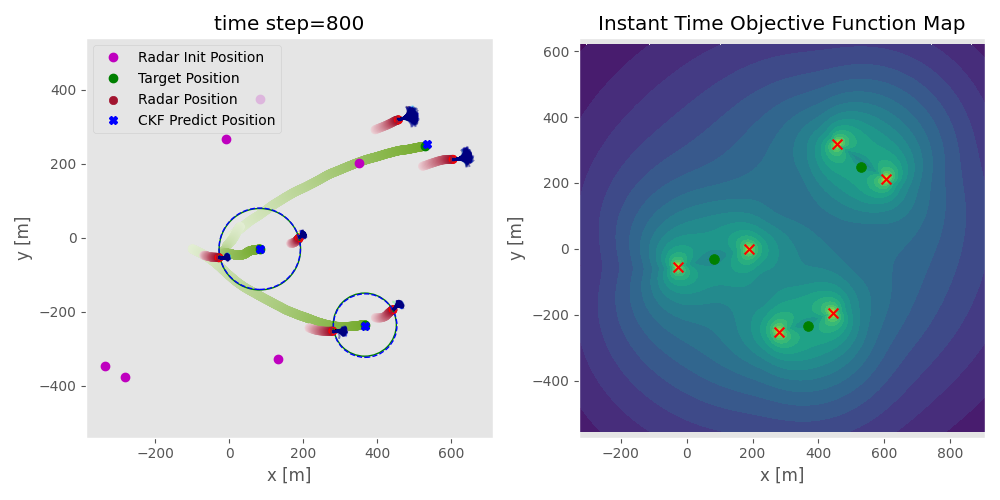}} \hfil 
\subfigure[Radars encircle the targets, but the upper most two radars move far away from
the upper most target.]{\includegraphics[width=8.5cm]{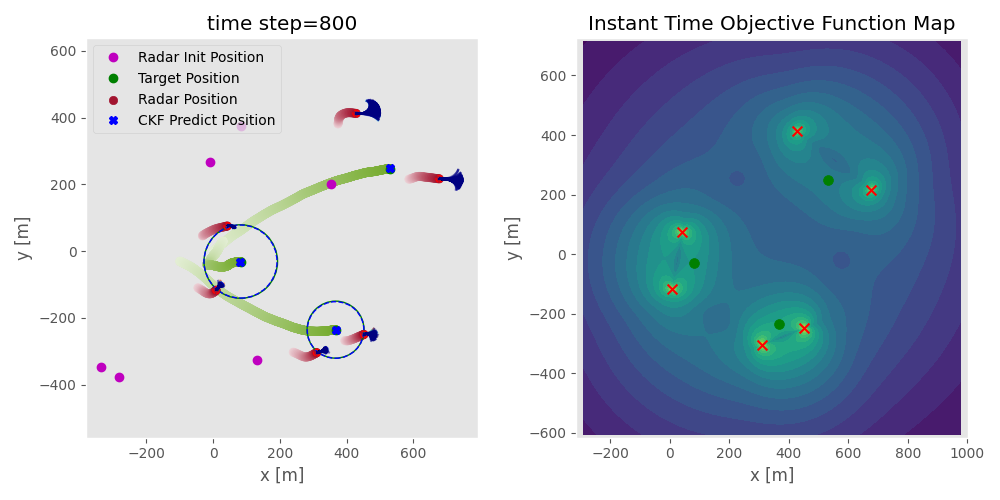}}\hfil  
\vspace{-.5em}
\subfigure[Two radars track each target individually by moving side-by-side parallel to the target's trajectory.]{\includegraphics[width=8.5cm]{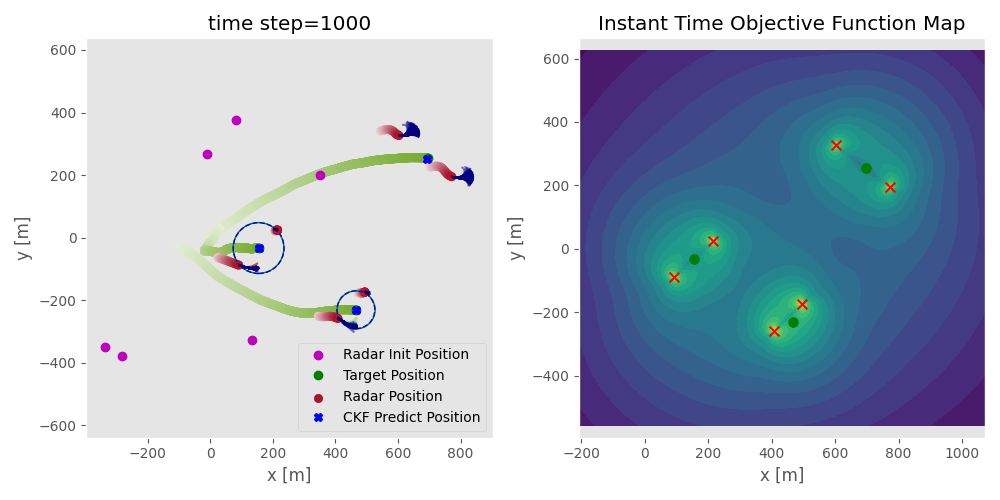}} \hfil 
\subfigure[Radars maintain encirclement formation around target, but the uppermost radar is far from the uppermost target]{\includegraphics[width=8.5cm]{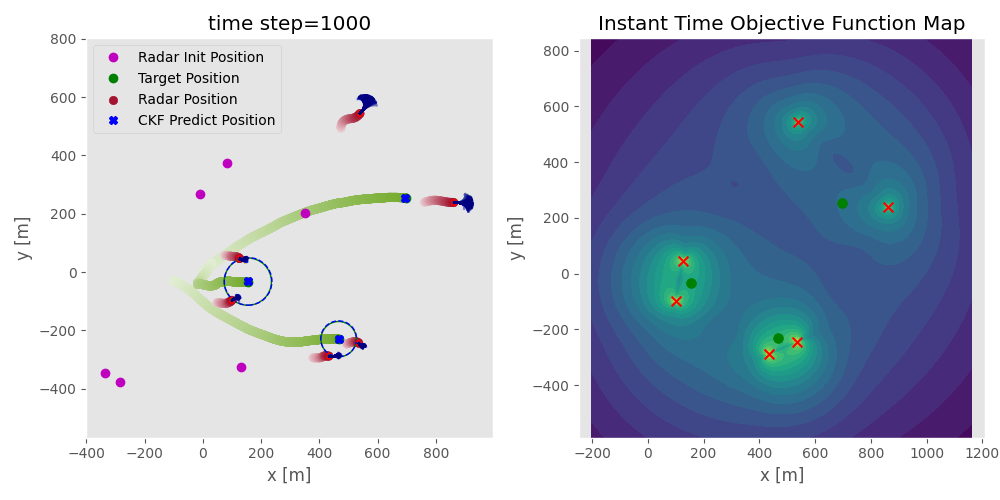}}\hfil  
\vspace{-.5em}
\caption{ \small  6 radars and 3 targets. The left column shows snapshots of target localization using a range measurement model with covariance dependent on radar-target distance \cite{godrich2010target}, while the right column depicts tracking with a range measurement model employing constant diagonal covariance \cite{bishop2010optimality}. In the first three rows, each subfigure's left panel shows a 3D projection onto 2D of the radar placement at a specific time. The colored lines emanating from the radars depict the weighted MPPI-planned trajectories, with blue indicating better paths and red indicating worse ones. Dashed red and green circles mark the radar's (10 [m]) and the target's (125 [m]) collision avoidance boundaries, respectively, while blue dashed circles represent the targets' collision avoidance boundary based on the CKF position estimate. The radar's trajectory is the last 25 time steps. The right panel displays the log determinant of the FIM (for the ddr measurement model) concerning the 2D target position for the specific time step, meshed over the free space area, and evaluated at each mesh point. }\label{appfig:noise_compare2}
\end{figure*}

\begin{figure*}[htp!]
\centering
\subfigure[Proposed approach: Pairs of radars move towards pairs of targets]{\includegraphics[width=8.5cm]{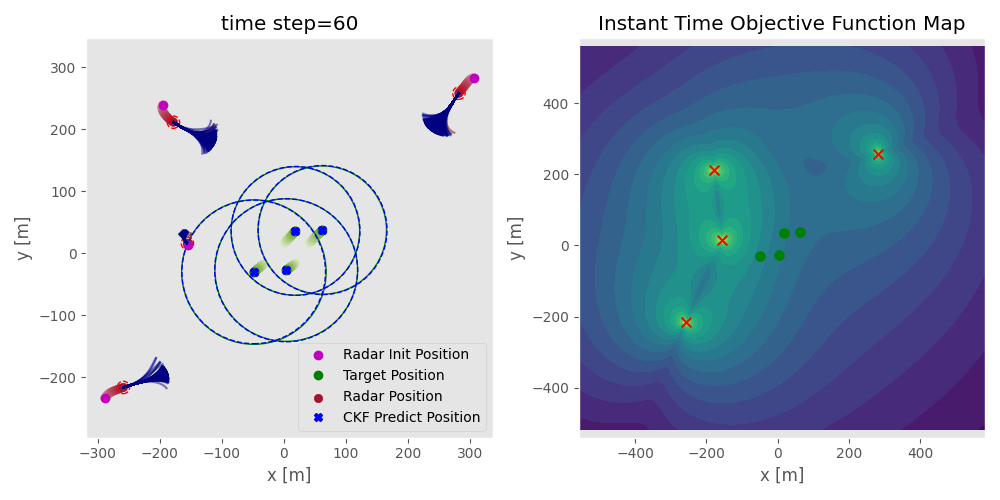}}\hfil
\subfigure[\cite{crasta2018multiple,bishop2010optimality,hung2020range} range model: the upperleft radar moves away from target]{\includegraphics[width=8.5cm]{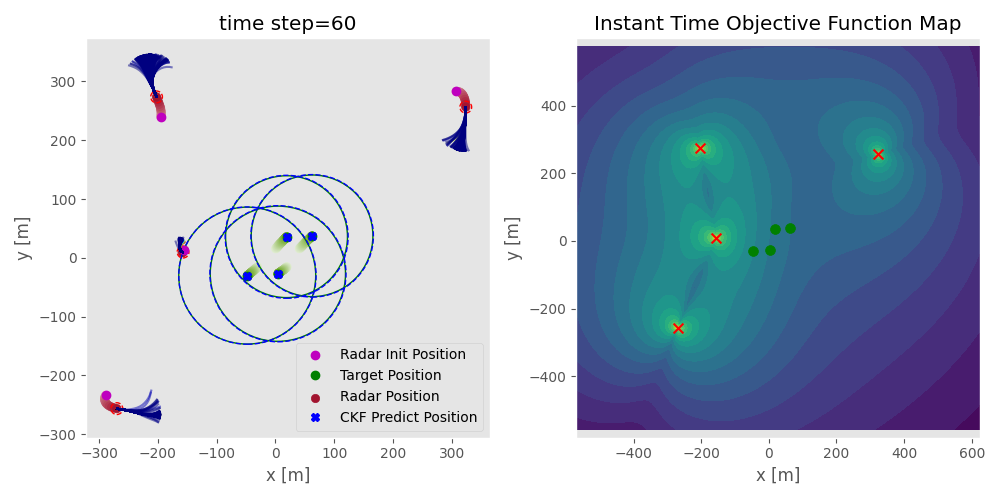}}\hfil 

\subfigure[Radars move toward targets as closely as feasible, with pairs of radars positioned side by side with pairs of targets.]{\includegraphics[width=8.5cm]{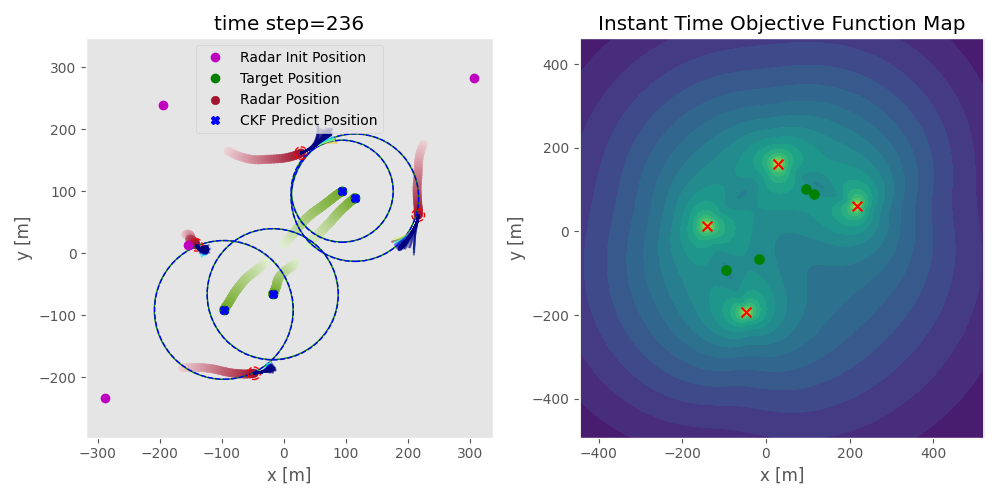}} \hfil 
\vspace{-.5em}
\subfigure[Radars encircle the targets, with the upper left radar positioned farther away from the targets.]{\includegraphics[width=8.5cm]{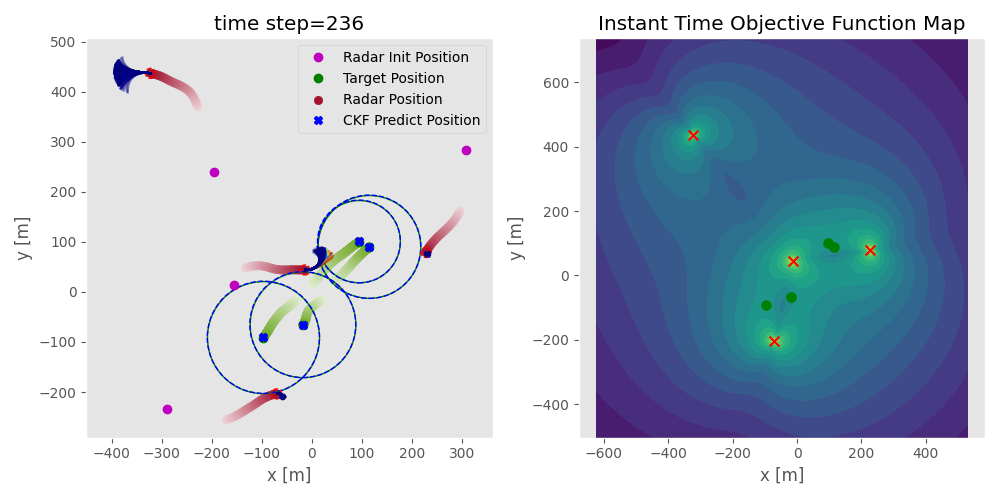}}\hfil  
\vspace{-.5em}
\subfigure[The pairs of targets are kept in the maximum information gain locations of the radar pairs.]{\includegraphics[width=8cm]{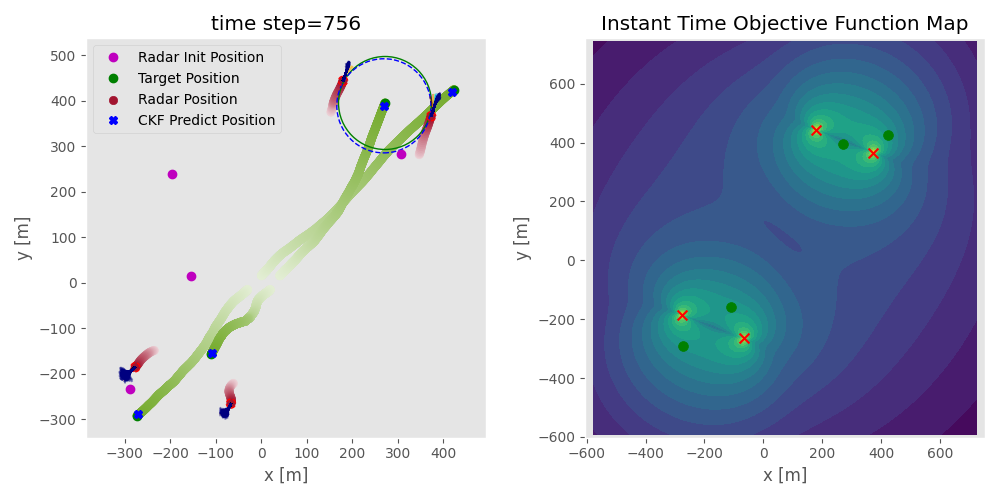}} \hfil 
\subfigure[The bottom left pair of targets only has one radar following, leading to worse target localization.]{\includegraphics[width=8.5cm]{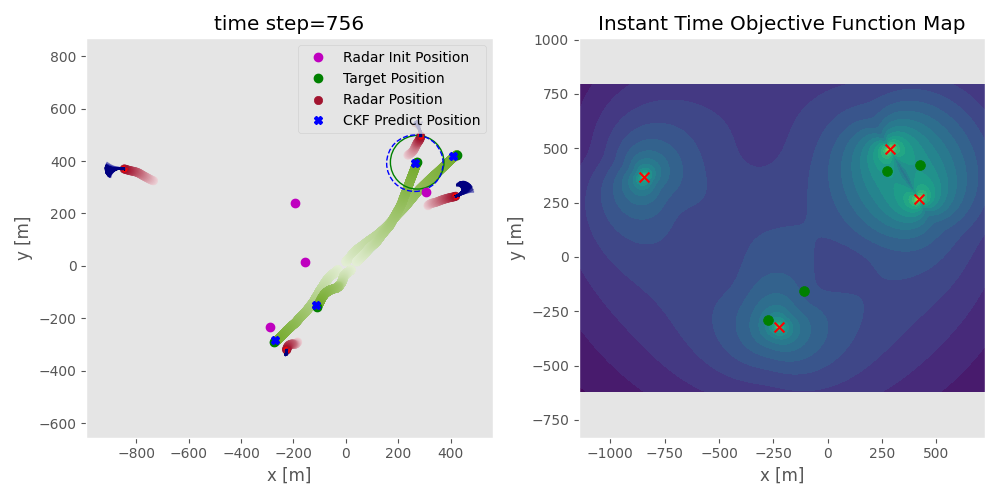}}\hfil  
\vspace{-.5em}
\subfigure[Pairs of radars track each pair of targets by moving side-by-side parallel to the pair of targets' trajectory.]{\includegraphics[width=8.5cm]{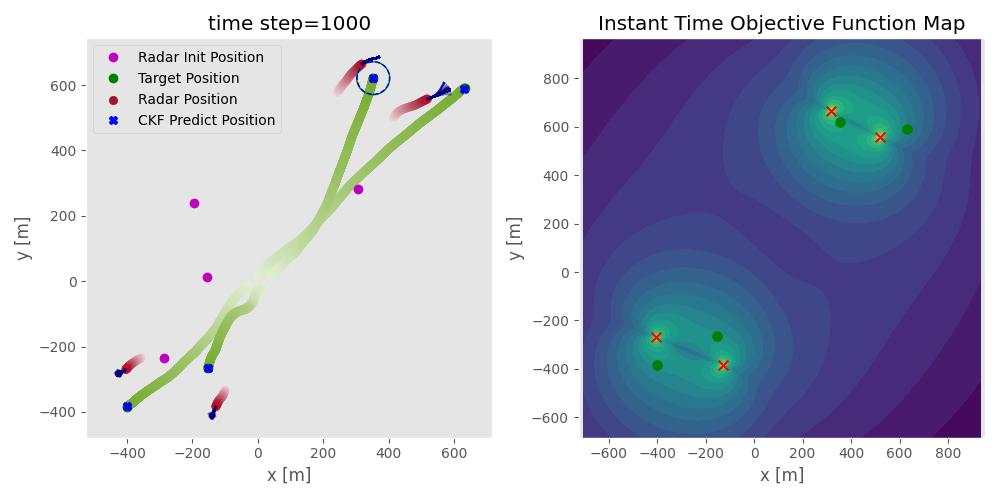}} \hfil 
\subfigure[Radars maintain encirclement formation around target, but the bottom left radar is far from the bottom left target]{\includegraphics[width=8.5cm]{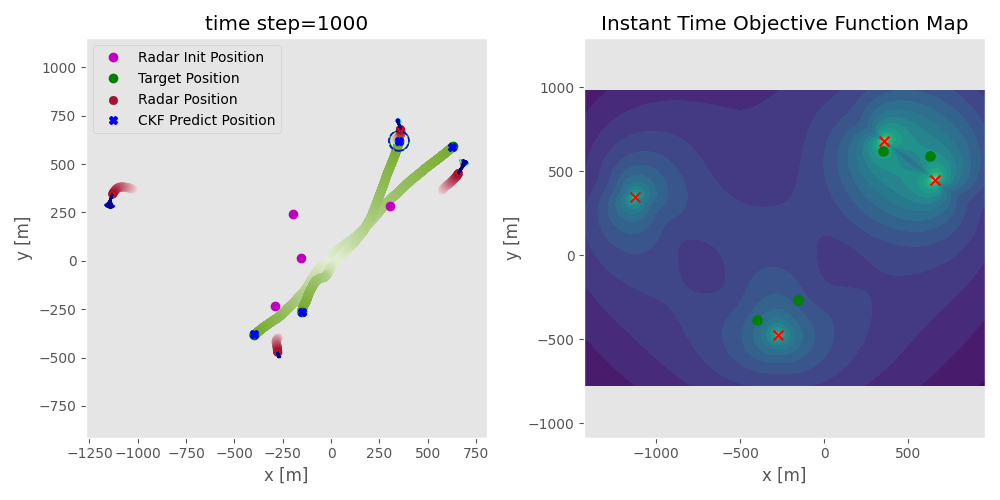}}\hfil  
\vspace{-.5em}
\caption{ \small 4 radars and 4 targets. The left column shows snapshots of target localization using a range measurement model with covariance dependent on radar-target distance \cite{godrich2010target}, while the right column depicts tracking with a range measurement model employing constant diagonal covariance \cite{bishop2010optimality}. In the first three rows, each subfigure's left panel shows a 3D projection onto 2D of the radar placement at a specific time. The colored lines emanating from the radars depict the weighted MPPI-planned trajectories, with blue indicating better paths and red indicating worse ones. Dashed red and green circles mark the radar's (10 [m]) and the target's (125 [m]) collision avoidance boundaries, respectively, while blue dashed circles represent the targets' collision avoidance boundary based on the CKF position estimate. The radar's trajectory is the last 25 time steps. The right panel displays the log determinant of the FIM (for the ddr measurement model) concerning the 2D target position for the specific time step, meshed over the free space area, and evaluated at each mesh point. }\label{appfig:noise_compare3}
\end{figure*}

The extended \gls{mc} simulation over 1000 time steps for 6 radars and 3 targets, and 4 radars and 4 targets are shown in \cref{appfig:noise_compare2} and \cref{appfig:noise_compare3},  respectively. GIFs for the undetermined, overdetermined, and full system are included in supplemental materials.

